\documentclass[a4paper,12pt]{article}

\pdfoutput=1

\usepackage{amsmath}
\usepackage{amssymb}
\usepackage{graphicx}
\usepackage{hyperref}
\usepackage{cite}
\usepackage{color}

\makeatletter
\@addtoreset{equation}{section}
\renewcommand{\theequation}{\thesection.\@arabic\c@equation}
\makeatother

\begin{document}

\begin{titlepage}

\vspace*{-15mm}   
\baselineskip 10pt   
\begin{flushright}   
\begin{tabular}{r}    
\end{tabular}   
\end{flushright}   
\baselineskip 24pt   
\vglue 10mm   

\begin{center}
{\Large\bf
 Quantum focusing conjecture and the Page curve
}

\vspace{8mm}   

\baselineskip 18pt   

\renewcommand{\thefootnote}{\fnsymbol{footnote}}

Yoshinori Matsuo\footnote[2]{ymatsuo@phys.kindai.ac.jp} 

\renewcommand{\thefootnote}{\arabic{footnote}}
 
\vspace{5mm}   

{\it  
 Department of Physics, Kindai University, \\Higashi-Osaka, Osaka 577-8502, Japan
}
  
\vspace{10mm}   

\end{center}

\begin{abstract}

The focusing theorem fails for evaporating black holes because 
the null energy condition is violated by quantum effects. 
The quantum focusing conjecture is proposed so that 
it is satisfied even if the null energy condition is violated. 
The conjecture states that the derivative of the sum of the area 
of a cross-section of the null geodesic congruence and 
the entanglement entropy of matters outside it is non-increasing. 
Naively, it is expected that the quantum focusing conjecture is violated 
after the Page time as both the area of the horizon and 
the entanglement entropy of the Hawking radiation are decreasing. 
We calculate the entanglement entropy after the Page time by using the island rule, 
and find the following results: 
(i) the page time is given by an approximately null surface, 
(ii) the entanglement entropy is increasing along the outgoing null geodesic even after the Page time, 
and (iii) the quantum focusing conjecture is not violated. 

\end{abstract}

\baselineskip 18pt   

\end{titlepage}

\newpage

\baselineskip 18pt

\tableofcontents


\section{Introduction}\label{sec:Intro}

The focusing theorem plays important roles in several studies in general relativity. 
The theorem states that the expansion of the congruence of null geodesics is non-increasing, 
\begin{equation}
 \frac{d\theta}{d \lambda} \leq 0 \ , 
\end{equation}
where $\theta$ is the expansion and $\lambda$ is an affine parameter. 
The focusing theorem is useful to prove universal structures of spacetime, 
and several important results in general relativity have been obtained. 
However, the focusing theorem is proven by using the null energy condition, 
which can be violated by quantum effects. 
Because of the violation of the null energy condition, 
many important results from the focusing theorem might not be true in real universe.  

One of the most important results of the focusing theorem is 
the second law of thermodynamics for the black hole entropy. 
The entropy of a black hole is given by \cite{Bekenstein:1972tm,Bekenstein:1973ur,Bekenstein:1974ax}
\begin{equation}
 S_\text{BH} = \frac{A}{4G_N} \ , 
 \label{SBH}
\end{equation}
where $A$ is the area of the event horizon and $G_N$ is the Newton constant. 
The event horizon is a null surface, and 
the expansion can be expressed in terms of the area as 
\begin{equation}
 \theta = \frac{1}{A} \frac{dA}{d \lambda} \ . 
\end{equation}
By using the focusing theorem, it can be shown that 
the expansion of the event horizon cannot be negative, 
and hence, the black hole entropy \eqref{SBH} satisfies 
the second law, 
\begin{equation}
 dS_\text{BH} \geq 0 \ . 
 \label{dSBH}
\end{equation}
However, if quantum effects are taken into account, the black hole loses its mass 
by emitting the Hawking radiation and evaporates eventually \cite{Hawking:1974sw,Hawking:1976ra}. 
The area of the horizon of evaporating black holes decreases with time, 
and the black hole entropy does not satisfy the second law. 
The black hole entropy decreases because of the emission of the Hawking radiation, 
and the entropy of the total system is expected to be non-decreasing. 
Bekenstein proposed the generalized second law which states that 
generalized entropy never decreases \cite{Bekenstein:1972tm,Bekenstein:1973ur,Bekenstein:1974ax}, 
\begin{equation}
 dS_\text{gen} \geq 0 \ , 
\end{equation}
where the generalized entropy $S_\text{gen}$ 
is the sum of the entropy of the black hole and 
the entropy outside the black hole, 
\begin{equation}
 S_\text{gen} = \frac{A}{4G_N} + S_\text{out} \ . 
 \label{Sgen}
\end{equation}

The generalized second law is originally proposed 
to solve the problem that the entropy of matters do not satisfy 
the second law when they fall into a black hole. 
The generalized second law can also be considered 
as a modification of \eqref{dSBH} so that 
it is satisfied even if quantum effects are taken into account. 
Then, it would be expected that the focusing theorem 
can be refined by using the generalized entropy, 
in a similar fashion to the generalized second law. 
In fact, the notion of the generalized entropy can be 
extended to more general surfaces \cite{Wall:2010jtc}. 

The quantum focusing conjecture \cite{Bousso:2015mna} 
states that the quantum expansion $\Theta$ is non-increasing, 
\begin{equation}
 \frac{d\Theta}{d \lambda} \leq 0 \ , 
 \label{QFC}
\end{equation}
where the quantum expansion is defined by replacing 
the area in the (classical) expansion by the generalized entropy \eqref{Sgen} as%
\footnote{%
To be more precise, the quantum focusing conjecture 
is originally proposed by using the local variation. 
In this paper, we focus on spherically symmetric surfaces 
and consider variation of the entire surface. 
} 
\begin{equation}
 \Theta = \frac{1}{A} \frac{d}{d \lambda} S_\text{gen} \ . 
 \label{QE}
\end{equation}
Here, the quantum expansion is defined on a codimension-2 surface 
which separates a Cauchy surface to two regions and is a cross-section of 
a null surface parameterized by the affine parameter $\lambda$, and  
$A$ is the area of the surface. 

In \cite{Bousso:2015mna}, the entropy of matters outside the surface, $S_\text{out}$, 
is defined by the von Neumann entropy, and hence, is nothing but the entanglement entropy.%
\footnote{%
The entanglement entropy has the area term as 
the most dominant term \cite{Bombelli:1986rw,Srednicki:1993im}. 
The area term diverges when the UV cutoff is taken away 
but is absorbed by the renormalization of the gravitational couplings \cite{Susskind:1994sm}. 
} 
For theories with gravity, the gravitational part of the theory gives an additional term 
to the entanglement entropy, 
\begin{equation}
 S_\text{grav} = \frac{A}{4G_N} \ , 
\end{equation}
for the (semi-)classical Einstein gravity, which is identical to the first term in \eqref{Sgen}. 
Thus, the entire of the generalized entropy \eqref{Sgen} should be interpreted as 
the entanglement entropy of the region outside the surface. 
It should be noted that the entanglement entropy of a union of two systems may not equal to 
but can be less than the sum of the entanglement entropy of each system. 
Thus, the generalized entropy cannot be interpreted as the entanglement entropy of the total system 
and the generalized second law might be violated. 

In fact, the quantum focusing conjecture would naively 
be expected to be violated for evaporating black holes after the Page time \cite{Page:1993wv,Page:2013dx}. 
In the generalized second law, the generalized entropy is non-decreasing 
because the entropy of matters increases when the entropy of the black hole is decreasing. 
If the Hawking radiation were purely thermal radiation, 
the entropy of the Hawking radiation would be increasing. 
However, the entanglement entropy of 
the Hawking radiation approximately equals to the black hole entropy after the Page time, 
\begin{equation}
 S_\text{out} \simeq \frac{A}{4 G_N} \ . 
\end{equation}
Then, the generalized entropy gives twice of the black hole entropy, 
\begin{equation}
 S_\text{gen} \simeq \frac{A}{2G_N} \ , 
\end{equation}
and is decreasing with time. 
Thus, the quantum focusing conjecture would naively be expected to be violated 
for outgoing null geodesics slightly outside the event horizon. 
It is already commented in \cite{Bousso:2015mna} that 
the quantum focusing conjecture may need to be modified for sufficiently old black holes, 
as it was considered the entanglement entropy after the Page time 
cannot be obtained in the semi-classical regime, at that moment. 

In this paper, we study the quantum focusing conjecture 
for evaporating black holes for more details. 
Recently, it has been proposed that the Page curve can be reproduced 
by using the island rule 
\cite{Penington:2019npb,Almheiri:2019psf,Almheiri:2019hni,Almheiri:2019yqk,Penington:2019kki,Almheiri:2019qdq}.%
\footnote{
See \cite{Almheiri:2020cfm} for a review and 
\cite{Akers:2019nfi,Chen:2019uhq,Almheiri:2019psy,Chen:2019iro,Akers:2019lzs,Liu:2020gnp,Marolf:2020xie,Balasubramanian:2020hfs,Bhattacharya:2020ymw,Verlinde:2020upt,Chen:2020wiq,Gautason:2020tmk,Anegawa:2020ezn,Hashimoto:2020cas,Sully:2020pza,Hartman:2020swn,Hollowood:2020cou,Krishnan:2020oun,Alishahiha:2020qza,Banks:2020zrt,Geng:2020qvw,Chen:2020uac,Chandrasekaran:2020qtn,Li:2020ceg,Bak:2020enw,Bousso:2020kmy,Anous:2020lka,Dong:2020uxp,Krishnan:2020fer,Hollowood:2020kvk,Engelhardt:2020qpv,Karlsson:2020uga,Chen:2020jvn,Chen:2020tes,Hartman:2020khs,Liu:2020jsv,Murdia:2020iac,Akers:2020pmf,Balasubramanian:2020xqf,Balasubramanian:2020coy,Sybesma:2020fxg,Stanford:2020wkf,Chen:2020hmv,Ling:2020laa,Marolf:2020rpm,Harlow:2020bee,Akal:2020ujg,Hernandez:2020nem,Chen:2020ojn,Matsuo:2020ypv,Goto:2020wnk,Hsin:2020mfa,Akal:2020twv,Colin-Ellerin:2020mva,KumarBasak:2020ams,Geng:2020fxl,Karananas:2020fwx,Wang:2021woy,Marolf:2021kjc,Bousso:2021sji,Geng:2021wcq,Ghosh:2021axl,Wang:2021mqq,Uhlemann:2021nhu,Kawabata:2021vyo,Akal:2021foz,Omiya:2021olc,Geng:2021hlu,Stanford:2021bhl,Balasubramanian:2021xcm,Akers:2021lms,Miyaji:2021lcq,Azarnia:2021uch,He:2021mst,Dong:2021oad,Shaghoulian:2021cef,Matsuo:2021mmi,Uhlemann:2021itz,Omidi:2021opl,Bhattacharya:2021nqj,Yu:2021rfg,Engelhardt:2022qts,Akers:2022max,Suzuki:2022xwv,Gan:2022jay,Rolph:2022csa,Bousso:2022tdb,Akers:2022qdl,Balasubramanian:2022gmo} 
for related works. 
} 
It provides a prescription to calculate the entanglement entropy 
which is consistent with unitarity, and 
the quantum focusing conjecture can be checked explicitly 
for a congruence of the outgoing null geodesics even after the Page time. 
We calculate the quantum expansion by using the island rule, 
and find the following results: 
\begin{enumerate}
\item[(i)]
The Page time is expressed by an approximately null surface and outgoing null geodesics 
cannot intersect the Page time. 

\item[(ii)]
Even after the Page time, the entanglement entropy is increasing along outgoing null geodesics, 
or equivalently, the quantum expansion of outgoing null geodesics is positive. 

\item[(iii)]
The quantum focusing conjecture is not violated even after the Page time. 

\end{enumerate}

This paper is organized as follows. 
In Sec.~\ref{sec:CFT}, we briefly review the focusing theorem and 
its failure near the evaporating black hole. 
In Sec.~\ref{sec:Island}, after a brief review on the island rule, 
we calculate the entanglement entropy to study the quantum focusing conjecture. 
Sec.~\ref{sec:Conclusion} is devoted for conclusion and discussions.


\section{Violation of classical focusing}\label{sec:CFT}


\subsection{Classical focusing theorem}\label{ssec:CFT}

We consider a congruence of null geodesics and the expansion is given by 
\begin{equation}
 \theta = \nabla_\mu k^\mu \ , 
\end{equation}
where $k^\mu$ is the tangent vector of the null geodesic which is normalized in terms of 
an affine parameter $\lambda$ as $dx^\mu = k^\mu d \lambda$. 
The expansion $\theta$ gives the fractional rate of change of area 
\begin{equation}
 \theta = \frac{1}{\delta A} \frac{d}{d \lambda} \delta A \ , 
\end{equation}
where $\delta A$ is an infinitesimal area of the purely transverse cross-section of the congruence. 

The Raychaudhuri equation for the null geodesic congruence gives 
\begin{equation}
 \frac{d \theta}{d \lambda} 
 = 
 - \frac{1}{2} \theta^2 - \sigma_{\mu\nu} \sigma^{\mu\nu} 
 - R_{\mu\nu} k^\mu k^\nu \ , 
 \label{Raychaudhuri}
\end{equation}
where $\sigma^{\mu\nu}$ is the shear tensor and $R_{\mu\nu}$ is the Ricci tensor. 
If the geometry is a solution of the Einstein equation and 
the null energy condition is satisfied, 
the last term in \eqref{Raychaudhuri} obeys 
\begin{equation}
 R_{\mu\nu} k^\mu k^\nu = 8 \pi G_N T_{\mu\nu} k^\mu k^\nu \geq 0 \ . 
\end{equation}
Therefore, the expansion is non-increasing, 
\begin{equation}
 \frac{d \theta}{d \lambda} \leq 0 \ . 
 \label{Focusing}
\end{equation}
In this paper, we call this condition as the classical focusing condition, 
to distinguish with the quantum focusing \eqref{QFC}. 

We consider spherically symmetric spacetimes 
and focus only on quantities independent of angular coordinates. 
The metric is expressed in general as 
\begin{equation}
 ds^2 = - C(u,v) du\,dv + r^2(u,v) d \Omega^2 \ , 
 \label{metric0}
\end{equation}
where $d \Omega^2$ is the metric on the unit 2-sphere. 
The expansion is now expressed in terms of the radius $r$ as 
\begin{equation}
 \theta = \frac{1}{4 \pi r^2} \frac{d}{d \lambda} \left(4 \pi r^2\right) = 2 \frac{d}{d \lambda} \log r \ . 
\end{equation}
Thus, the condition \eqref{Focusing} implies that the null geodesic cannot reach $r\to\infty$ 
once the expansion becomes negative at some point on the geodesic. 
In other words, the expansion is always positive on 
all outgoing null geodesics which goes to future null infinity 
if $r\to \infty$ at the future null infinity.


\subsection{Geometry around an evaporating black hole}\label{ssec:Vaidya}

We consider a Schwarzschild black hole which evaporates by emitting the Hawking radiation. 
If effects of the Hawking radiation are ignored, the geometry is given by the Schwarzschild solution, 
and then, the classical focusing condition \eqref{Focusing} is satisfied. 
However, the ingoing negative energy appears associated to the Hawking radiation. 
The geometry is affected by the ingoing negative energy 
and the classical focusing condition is violated. 

We consider the spherically symmetric spacetime and the metric is given by \eqref{metric0}. 
We focus only on degrees of freedom which are independent of angular coordinates, 
or equivalently, apply the s-wave approximation. 
Then, massless fields around the black hole can be approximated by two-dimensional fields 
which is obtained by ignoring angular directions. 
The energy-momentum tensor in four dimensional spacetime is given in terms of 
that of two dimensional fields as 
\begin{equation}
 T_{\mu\nu} = \frac{1}{4\pi r^2} T_{\mu\nu}^{(2D)} \ , 
\end{equation}
for temporal and radial directions, and the other (angular) components vanish. 
The energy-momentum tensor of two-dimensional massless fields is 
determined up to the integration constants by the conservation law 
\begin{equation}
 \nabla_\mu T^{(2D)\ \mu}{}_\nu = 0 \ , 
\end{equation}
and the trace anomaly, 
\begin{equation}
 T^\mu{}_{\mu} = \frac{c}{24\pi} R^{(2D)} \ , 
\end{equation}
where $R^{(2D)}$ is the two-dimensional Ricci scalar and $c$ is the central charge. 
Then, the energy-momentum tensor is obtained as \cite{Davies:1976ei,Davies:1976hi} 
\begin{align}
T_{uu} &=
- \frac{c}{48\pi^2 r^2} C^{1/2} \partial_u^2 C^{-1/2} + \frac{c\,F(u)}{96\pi^2 r^2} \ , 
\label{Tuu-vac}
\\
T_{vv} &=
- \frac{c}{48\pi^2 r^2} C^{1/2} \partial_v^2 C^{-1/2} + \frac{c\,\bar F(v)}{96\pi^2 r^2} \ , 
\label{Tvv-vac}
\\
T_{uv} &=
- \frac{c}{96\pi^2 r^2 C^2}
\left[ C \partial_u\partial_v C - \partial_u C \partial_v C \right],
\label{Tuv-vac}
\end{align}
where $F(u)$ and $\bar F(v)$ are the integration constants, and $C$ is the metric defined in \eqref{metric0}. 

\begin{figure}[t]
\begin{center}
\includegraphics[scale=0.5]{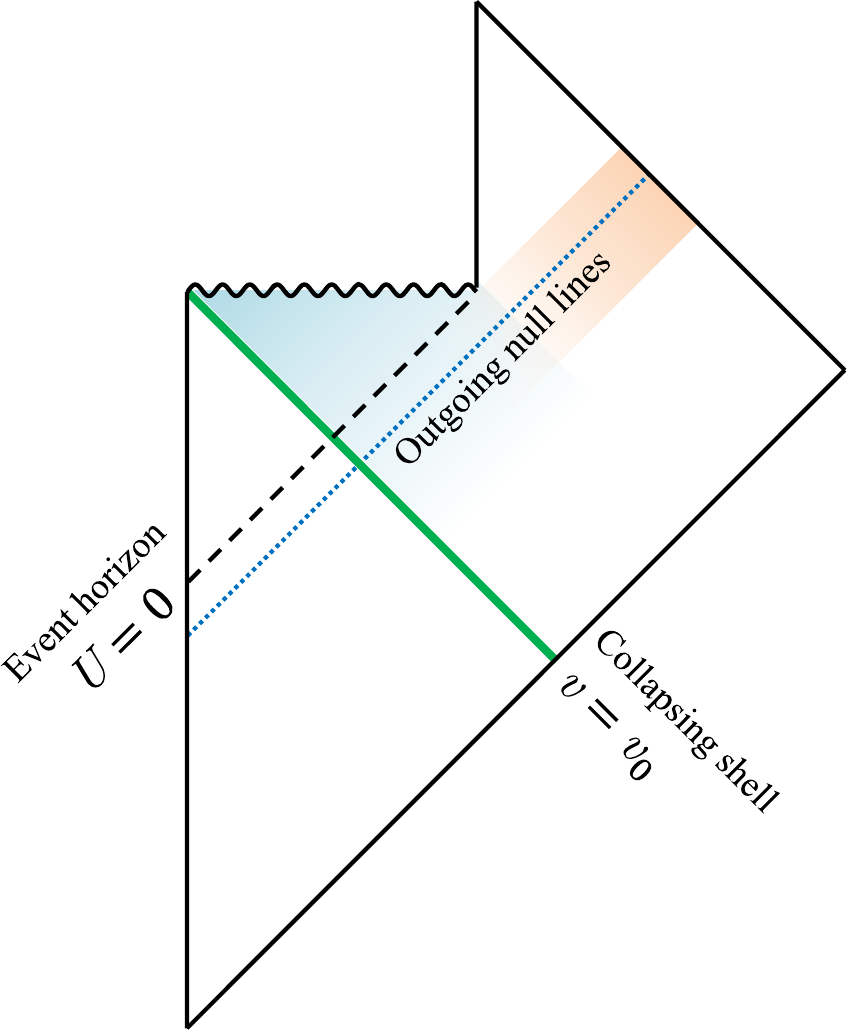}
\caption{%
The Penrose diagram of the evaporating black hole geometry. 
The black hole is formed by the gravitational collapse of a thin shell at $v=v_0$, 
which is indicated by the thick line. 
The dashed line and wavy line are the event horizon (at $U=0$) and singularity, respectively. 
We consider the expansion of outgoing null lines near the horizon, 
which is indicated by the dotted line. 
Quantum effects induces the incoming negative energy near the horizon 
and outgoing (Hawking) radiation away from the shell, 
which are indicated by shaded regions. 
}\label{fig:Penrose}
\end{center}
\end{figure}

For the Schwarzschild spacetime, the advanced time $v$ and retarded time $u$ can be chosen as 
\begin{align}
 dv &= dt + \frac{dr}{1 - \frac{r_h}{r}} \ , 
&
 du &= dt - \frac{dr}{1 - \frac{r_h}{r}} \ , 
\end{align}
where $r_h$ is the Schwarzschild radius, and then, $C$ is given by 
\begin{equation}
 C = 1 - \frac{r_h}{r} \ . 
\end{equation}
For simplicity, we consider a black hole which is formed by the gravitational collapse 
of a null thin shell at $v=v_0$ (Fig.~\ref{fig:Penrose}), 
and estimate the energy-momentum tensor on the Schwarzschild spacetime. 
The incoming energy $T_{vv}$ must be zero in the past null infinity, 
and then, approximately given near the (future) horizon $r\sim r_h$ as 
\begin{equation}
 T_{vv} \simeq - \frac{c}{768\pi^2 r_h^4} \ . 
 \label{Tvv}
\end{equation}
The outgoing energy $T_{uu}$ must be zero inside the collapsing shell. 
Since the shell is collapsing at the speed of light, 
the tangential pressure on the collapsing shell must be zero. 
Then, $T_{uu}$ cannot have discontinuity at the shell, 
and then, vanishes near the future horizon, 
\begin{equation}
 T_{uu} \simeq 0 \ . 
 \label{Tuu}
\end{equation}
Thus, the ingoing energy becomes negative, and the outgoing energy becomes zero near the horizon. 

A solution of the Einstein equation with the incoming energy is given by the ingoing Vaidya metric, 
\begin{equation}
 ds^2 = - \left(1-\frac{r_h(v)}{r}\right) dv^2 + 2 dv\,dr + r^2 d \Omega^2 \ , 
 \label{Vaidya}
\end{equation}
and then, the energy-momentum tensor has non-zero component only in 
\begin{equation}
 T_{vv} = \frac{\dot r_h(v)}{8\pi G_N r^2} \ , 
\end{equation}
where $\dot r_h = \frac{dr_h}{dv}$, and all the other components vanish. 
Thus, the geometry near the future horizon can be approximated by 
the ingoing Vaidya metric with the negative incoming energy \eqref{Tvv}, or equivalently,  
\begin{equation}
 \dot r_h(v) = - \frac{c\,G_N}{96\pi r_h^2(v)} \ . 
 \label{drh}
\end{equation}


\subsection{Violation of classical focusing condition}\label{ssec:violation}

The locus $r(v)$ of outgoing null lines in the ingoing Vaidya spacetime 
is given by the solution of 
\begin{equation}
 \frac{dr}{dv} = \frac{1}{2} \left(1-\frac{r_h(v)}{r}\right) \ . 
 \label{locus}
\end{equation}
The ingoing null lines, $v = \mathrm{const.}$, are simply parameterized by $r$, 
and hence, the radius $r$ is always decreasing with time along the ingoing null lines. 
Eq.~\eqref{locus} implies that $r$ is increasing along the outgoing null lines 
for $r>r_h(v)$ and decreasing for $r<r_h(v)$. 
Thus, $r_h(v)$ is nothing but the apparent horizon, 
which equals to the Schwarzschild radius up to $\mathcal O(G_N)$ corrections. 
In this paper, we focus on black holes which is much larger than the Planck length $r_h^2(v)\gg G_N$.  
Since the change of the apparent horizon \eqref{drh} is as small as $\mathcal O(G_N)$, 
eq.~\eqref{locus} is solved to the linear order of the small-$G_N$ expansion as 
\begin{equation}
 r(v) 
 = 
 r_h(v) \left(1 + W(X)\right) 
 + r_h(v) \dot r_h(v) 
 \frac{2 + 4 W(X) - W^3(X)}{1 + W(X)} 
 + \mathcal O(G_N^2) 
 \ , 
 \label{r-full}
\end{equation}
where $W(x)$ is the Lambert W function which is defined by $x = W(x) e^{W(x)}$, and 
\begin{equation}
 X = -\frac{r_h^2(v_0)}{2r_h^3(v)}\,U \exp\left[\int_{v_0}^v \frac{dv'}{2r_h(v')} \right] \ , 
\end{equation}
where $v=v_0$ is the position of the ingoing null shell, and $U$ is an integration constant. 

Near the apparent horizon, the radius $r(v)$ is approximated as 
\begin{equation}
 r(v) 
 \simeq 
 r_h(v) - \frac{r_h^2(v_0)}{2r_h^2(v)}\,U \exp\left[\int_{v_0}^v \frac{dv'}{2r_h(v')} \right] 
 + 2 r_h(v) \dot r_h(v) \ . 
 \label{r}
\end{equation}
For $U>0$, the radius on the outgoing null lines is decreasing. 
For $U < 0$, the second term in \eqref{r} is monotonically increasing 
but the first term is decreasing for the ingoing Vaidya metric. 
By using \eqref{drh}, the expansion $\theta$ is estimated as 
\begin{equation}
 \theta 
 \propto 
 \frac{\partial r}{\partial v} 
 \simeq 
 - \frac{c\,G_N}{96\pi r_h^2(v)} 
 - \frac{r_h^2(v_0)}{4r_h^3(v)} U 
 \exp\left[\int_{v_0}^v \frac{dv'}{2r_h(v')}\right] \ . 
\end{equation}
The radius $r$ is always increasing only for $U < - \frac{c G_N}{24\pi r_h(v)}$. 
For $0 > U > - \frac{c G_N}{24\pi r_h(v)}$, 
$r$ is increasing for sufficiently large $v$ but decreasing for $v\gtrsim v_0$. 
Therefore, the expansion $\theta$ is increasing at some point and hence 
the classical focusing condition \eqref{Focusing} is violated. 
The violation of \eqref{Focusing} can also be seen from 
the relative position between the apparent horizon and event horizon. 
The outgoing null lines will eventually reach the singularity for $U>0$ 
but go to the future null infinity for $U<0$. 
Thus the event horizon is the null surface at $U=0$, and the radius is given by 
\begin{equation}
 r = r_\text{EH}(v) \equiv r_h(v) + 2 r_h(v) \dot r_h(v) + \mathcal O(G_N^2) \ , 
\end{equation}
which is smaller than $r_h(v)$ for given $v$, or equivalently on an incoming null line. 
Since the radius $r$ is monotonically decreasing along incoming null lines 
in the ingoing Vaidya spacetime, 
the event horizon is inside the apparent horizon, 
The expansion is negative outside the event horizon, 
and hence, the classical focusing condition is violated. 

To see the violation of the classical focusing condition more explicitly, 
we introduce an affine parameter $\lambda$ as 
\begin{equation}
 d \lambda = - \frac{\partial r(v,U)}{\partial U}\,dv \ , 
\end{equation}
which is integrated as 
\begin{equation}
 \lambda 
 = 
 - \frac{r_h^2(v_0)}{r_h(v)}\,\exp\left[\int_{v_0}^v \frac{dv'}{2r_h(v')} \right] 
 - \frac{r_h^4(v_0)}{2r_h^4(v)}\,U \exp\left[\int_{v_0}^v \frac{dv'}{r_h(v')} \right] 
 + \mathcal O(G_N) \ , 
\end{equation}
near the apparent horizon. 
By using \eqref{drh}, the derivative of the expansion is estimated as 
\begin{equation}
 \frac{d \theta}{d \lambda} 
 = 
 - \frac{U^2}{2 r_h^4(v)} 
 + \frac{c\,G_N}{24\pi r_h^4(v_0)} \exp\left[-\int_{v_0}^v \frac{dv}{r_h(v)}\right] 
 \ . 
\end{equation}
Although the expansion $\theta$ becomes negative inside the apparent horizon, 
which is located within a distance of order of the Planck length,%
\footnote{%
The proper distance $L$ from the event horizon in the Schwarzschild spacetime 
is estimated as $L \simeq \sqrt{r_h(r-r_h)}$ in the vicinity of the horizon. 
} 
\begin{equation}
 r(v) - r_\text{EH}(v) 
 = 
 - \frac{r_h^2(v_0)}{2 r_h^2(v)} U \exp\left[\int_{v_0}^v \frac{dv}{2r_h(v)}\right] 
 < 
 \frac{c\,G_N}{48\pi r_h(v)} \ , 
 \label{AH}
\end{equation}
the classical focusing condition is violated in a wider region, 
\begin{equation}
 r(v) - r_\text{EH}(v) 
 < 
 \sqrt{\frac{c\,G_N}{48\pi}} \ , 
 \label{FTF}
\end{equation}
which is no longer of order of the Planck length from the event horizon. 

The integration constant $U$ distinguishes different null lines and 
can be used as the retarded time. 
In terms of $(v,U)$ coordinates, the metric is expressed as 
\begin{equation}
 ds^2 = 2 \frac{\partial r(v,U)}{\partial U}\,dU\,dv + r^2(v,U) d \Omega^2 \ . 
 \label{metric}
\end{equation}
Near the apparent horizon, the radius is approximated as \eqref{r}, 
and then, the metric becomes 
\begin{equation}
 ds^2 
 = 
 - \frac{r_h^2(v_0)}{r_h^2(v)} \exp\left[\int_{v_0}^v \frac{dv'}{2r_h(v')} \right] dU\,dv 
 + r^2(v,U) d \Omega^2 \ .  
\end{equation}
The ingoing Vaidya spacetime is connected to the flat spacetime at $v=v_0$, 
and the junction condition is obtained from the relation to the radius. 
We have defined the integration constant $U$ so that it satisfies 
the same relation to that in the flat spacetime 
\begin{equation}
 dU = 2 dr \ , 
\end{equation}
near the apparent horizon. 
Therefore, the retarded coordinate $U$ near the apparent horizon 
is identical to that in the flat spacetime, where the metric in the flat spacetime is given by 
\begin{equation}
 ds^2 = - dU\,dv \ . 
 \label{flat}
\end{equation}
Away from the horizon, the retarded time in the flat spacetime is given in terms of \eqref{r-full} as 
\begin{equation}
 U_\text{flat} = 2 r(v_0,U) + \text{const}. 
\end{equation}


\section{Quantum focusing after the Page time}\label{sec:Island}


\subsection{Entanglement entropy and islands}\label{ssec:island}

In this paper, we study the quantum focusing conjecture 
in an evaporating black hole geometry. 
We consider the Einstein gravity with massless matter fields in four dimensions. 
For simplicity, we restrict the massless fields to s-waves, 
which have no angular momentum and can be described effectively by massless fields in two dimensions. 
To be more explicit, the action of the model is given by 
the Einstein-Hilbert action with the Gibbons-Hawking term, 
\begin{align}
 \mathcal I &= \mathcal I_\text{grav} + \mathcal I_\text{matter} \ , 
\\
 \mathcal I_\text{grav} 
 &= 
 \frac{1}{16\pi G_N} \int_{\mathcal M} d^4x \sqrt{-g}\,R 
 + \frac{1}{8\pi G_N} \int_{\partial\mathcal M} d^3x \sqrt{-h}\,K \ . 
\end{align}
Here, we focus only on the spherically symmetric configurations and 
the metric can be expressed as \eqref{metric0}. 
For massless scalar fields, the action is approximated by the two-dimensional action, 
\begin{align}
 \mathcal I_\text{matter} 
 &= 
 \frac{1}{2} \sum_{i=1}^c \int d^2x \sqrt{-g_\text{(2D)}} \, 
 g^{\mu\nu}_\text{(2D)} \partial_\mu \phi^i \partial_\nu \phi^i 
 = 
 \frac{1}{2} \sum_{i=1}^c \int du\,dv\,\partial_u \phi^i \partial_v \phi^i  
 \ , 
\end{align}
where $\phi^i$ is a two-dimensional scalar, or equivalently, 
the s-wave of a four-dimensional scalar which has no angular dependence, 
\begin{equation}
 \phi^i_\text{(4D)} = \frac{1}{\sqrt{4\pi}\,r}\,\phi^i \ . 
\end{equation}
We mainly consider the quantum focusing conjecture in this model, 
and then, discuss how modes with non-zero angular momenta affect our result, later. 

As we have discussed in the previous section, 
the generalized entropy of the quantum focusing conjecture 
is defined by using the von Neumann entropy, and hence, 
identical to the entanglement entropy outside the cross-section of the null surface. 
Here, we consider the spherically symmetric cross-section 
on which coordinates $(v,U)$ in \eqref{metric} are constant. 
In this paper, we refer to the cross-section and the region outside it as $b$ and $R$, respectively. 
The entanglement entropy of the region $R$ is given by the island rule 
\cite{Penington:2019npb,Almheiri:2019psf,Almheiri:2019hni,Almheiri:2019yqk,Penington:2019kki,Almheiri:2019qdq}, 
\begin{equation}
 S(R) 
 = 
 \min\left\{\mathrm{ext}\left[
 \sum_{\partial R, \partial I} \frac{A}{4G_N} + S_\text{matter}^\text{(non-local)}
 \right]\right\} \ .  
 \label{S(R)}
\end{equation}
We apply the s-wave approximation, 
and the matters can be described by two-dimensional massless fields. 
The entanglement entropy of matters $S_\text{matter}^\text{(non-local)}$ is given by 
\begin{align}
 S_\text{matter}^\text{(non-local)} 
 &= 
 \frac{c}{6} \sum_{i,j}\log\left|x_i-x'_j\right|^2 
 - \frac{c}{6} \sum_{i<j}\log\left|x_i-x_j\right|^2 
 - \frac{c}{6} \sum_{i<j}\log\left|x'_i-x'_j\right|^2 
\notag\\&\quad 
 + \frac{c}{12} \sum_i \log\left|g_{uv}(x_i)g_{uv}(x'_i)\right| \ . 
 \label{Smatter}
\end{align}
where $(x_i,x'_i)$ are the endpoints (boundaries) of the region, 
and the distance $\left|x_i - x_j\right|$ is expressed in terms of the null coordinates $(u,v)$ as 
\begin{equation}
 \left|x_i-x_j\right|^2 = \left|\left(u_i-u_j\right)\left(v_i-v_j\right)\right| \ , 
 \label{distance}
\end{equation}
and similarly for $\left|x_i - x'_j\right|$ and $\left|x'_i - x'_j\right|$. 
The distance \eqref{distance} is not the proper distance in general, 
but should be given in terms of the coordinates 
which is used to define the annihilation and creation operators for the vacuum state 
(For more details, see appendix~\ref{app:entropy}). 
For the vacuum state $T_{\mu\nu} = 0$ in the flat spacetime, 
the null coordinates should be chosen so that $C=1$ in \eqref{metric0}. 

According to the island rule \eqref{S(R)}, the entanglement entropy of the region $R$ 
may effectively contain contributions from the regions called islands. 
The boundaries of islands are given by the quantum extremal surfaces 
\cite{Ryu:2006bv,Hubeny:2007xt,Lewkowycz:2013nqa,Faulkner:2013ana,Engelhardt:2014gca,Dong:2016hjy,Dong:2017xht}, 
which extremize the entanglement entropy. 
In all configurations of the quantum extremal surfaces, 
including the configurations without islands, 
the configuration with the minimum entanglement entropy is realized.%
\footnote{%
To be more precise, all configurations of the quantum extremal surfaces 
are saddle points of the path integral on the replica geometries, 
and the entanglement entropy is minimized for the maximum of the partition function. 
Thus, the entanglement entropy is approximated by 
the minimum entanglement entropy in the saddle points. 
} 
For our setup of the region $R$, the island is a connected region and 
cannot be a union of disconnected regions. 
We refer to the island and the quantum extremal surface as $I$ and $a$, respectively. 
The area term in \eqref{S(R)} is the sum over all boundaries of $R$ and $I$, 
and the sums in \eqref{Smatter} are over all endpoints of $R$ and $I$. 
Note that we do not consider the quantum focusing of the quantum extremal surface $a$, in this paper. 
We consider the quantum focusing of $b$, which is the (inner) boundary of $R$, 
and the island appears associated to $b$. 
Thus, $b$ moves along the outgoing null surface but $a$ may not. 

In this paper, we calculate the entanglement entropy of the region $R$ in the evaporating black hole geometry.%
\footnote{%
The island rule in the same geometry is already discussed 
very briefly in Appendix~B of \cite{Matsuo:2020ypv}, 
but we explore more details to study the quantum focusing conjecture, here. 
} 
As is discussed in Sec.~\ref{ssec:Vaidya}, we consider a black hole formed by 
a gravitational collapse of a thin shell for simplicity, 
but it is straightforward to generalize the result for arbitrary collapses. 
The spacetime after the collapse of the shell is approximated by 
the ingoing Vaidya metric with the negative incoming energy. 
Thus, the geometry of the evaporating black hole is obtained by 
connecting the ingoing Vaidya spacetime and flat spacetime at the locus of the shell, 
namely $v=v_0$ in $v$ coordinates of \eqref{Vaidya}, \eqref{metric} and \eqref{flat}. 
As we discussed in Sec.~\ref{ssec:Vaidya}, 
the incoming energy is zero in the past null infinity, 
and the outgoing energy is zero before the gravitational collapse of the shell. 
The spacetime is flat both in the past null infinity and before the gravitational collapse. 
Thus, the null coordinates in the expression \eqref{Smatter} and \eqref{distance} 
should be the advanced time in the past null infinity and 
the retarded time before the gravitational collapse.%
\footnote{%
For appropriate choice of the coordinates for the entanglement entropy formula 
in the case of evaporating black holes, 
see also \cite{Penington:2019npb,Gautason:2020tmk,Matsuo:2020ypv}.
} 
Near the horizon of the ingoing Vaidya metric, 
these null coordinates are identical to the coordinates $(v,U)$ in \eqref{metric}. 

\begin{figure}[t]
\begin{center}
\includegraphics[scale=0.5]{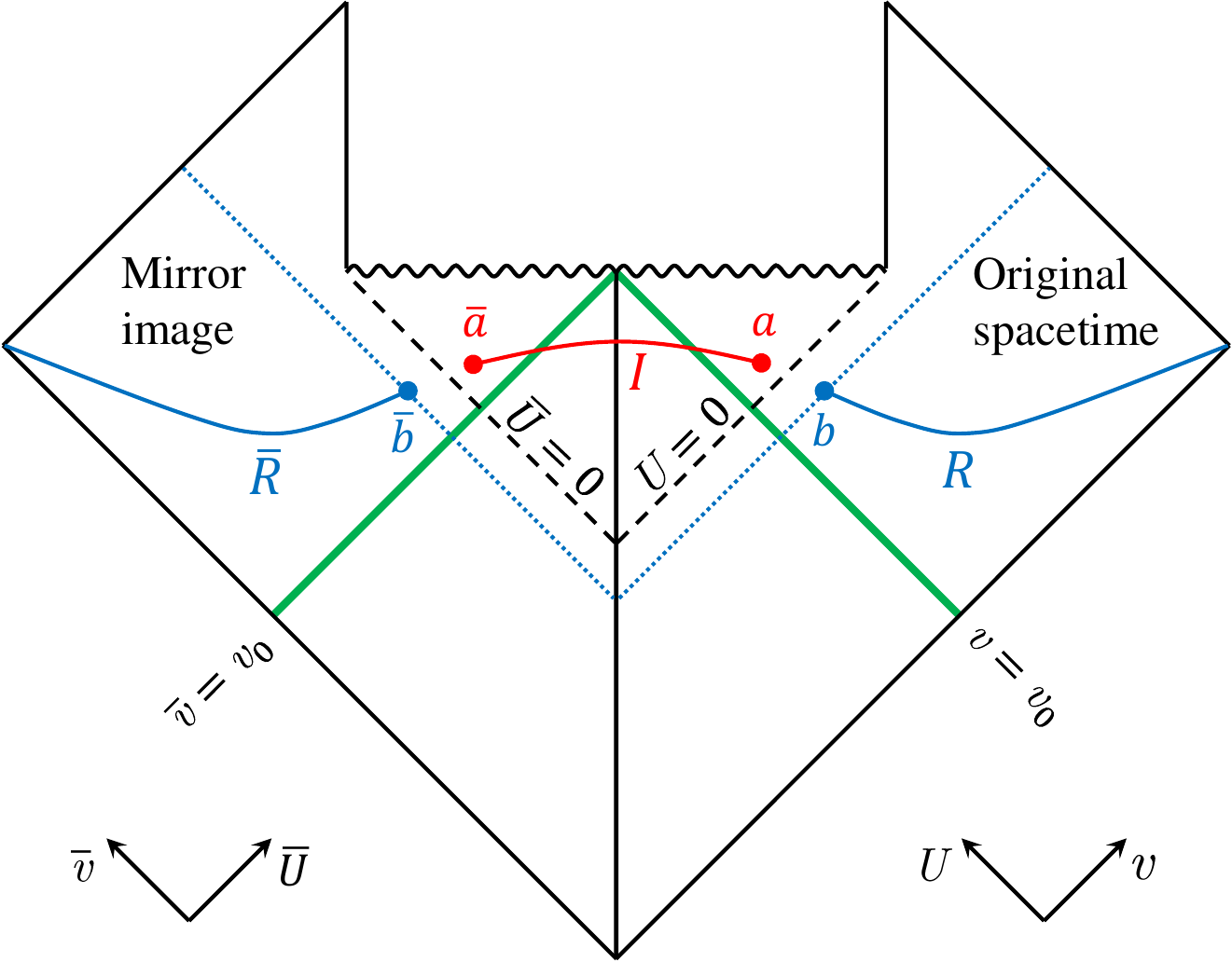}
\caption{%
The Penrose diagram of the evaporating black hole spacetime and its mirror image. 
We consider the quantum expansion of the outgoing null surface and 
the cross-section of the null surface is $b$. 
The codimension-2 surface $b$ separates the timeslice to its inside and outside, 
and $R$ is the region outside $b$. 
Since incoming modes turn into outgoing modes after passing through the origin $r=0$, 
we introduce the mirror image on the other side of $r=0$. 
The mirror image of $b$ and $R$ are referred to as $\bar b$ and $\bar R$. 
We calculate the entanglement entropy of the region $R\cup \bar R$, 
which gives twice of the entanglement entropy of $R$ due to the double counting of each modes. 
}\label{fig:mirror}
\end{center}
\end{figure}

Now, we evaluate the entanglement entropy by using the island rule. 
We first consider the configuration without islands. 
Ingoing modes turn into outgoing modes by passing through the origin $r=0$, 
and hence, the origin $r=0$ should be treated as a boundary with the totally reflecting boundary condition. 
We introduce the mirror image (See Fig.~\ref{fig:mirror}), and then, the formula \eqref{Smatter} gives 
\begin{equation}
 S_\text{matter}^\text{(non-local)}
 = 
 \frac{c}{12} \log\left|g_{uv}(x_b)\left(v_b - v_b'\right)\left(U_b-U_b'\right)\right| \ , 
 \label{NoIsland0}
\end{equation}
where $x_b = (v_b, U_b)$ stands for the position of $b$ (the inner boundary of $R$), 
and $(v'_b, U'_b)$ is the position of the mirror image of $b$. 
Note that the entanglement entropy is a half of \eqref{Smatter}, 
since the entanglement entropy becomes twice when the mirror image is introduced. 
The metric is given by $g_{uv} = - \frac{1}{2}$ before the collapse, $v_b < v_0$, 
and approximated as 
\begin{equation}
 g_{uv} = \frac{\partial r(v,U)}{\partial U} 
 \simeq - \frac{r_h^2(v_0)}{2r_h^2(v)} \exp\left[\int_{v_0}^v \frac{dv'}{2 r_h(v')}\right] \ , 
\end{equation}
near the horizon $r(v_b,U_b) - r_\text{EH}(v_b) \ll r_h(v_b)$ after the collapse $v_b > v_0$. 

We refer to the mirror image of the coordinates $v$ and $U$ as $\bar v$ and $\bar U$. 
The surface $b$ is located at $(v,U) = (v_b,U_b)$. 
In $(\bar v,\bar U)$-coordinates, 
the mirror image of $b$ is located at $(\bar v,\bar U) = (v_b,U_b)$, 
which is identical to $(v,U) = (v'_b,U'_b)$. 
The mirror image of the event horizon is located at $\bar U=0$, and 
the mirror image of the collapsing shell is at $\bar v = v_0$. 
Then, the event horizon at $r=0$ is expressed in these coordinates as 
\begin{align}
 v &= \bar v = v_0 - r_h(v_0) + \mathcal O(G_N) \ , 
 & 
 U &= \bar U = 0 \ . 
\end{align}
Since the advanced and retarded time are exchanged with each other in the mirror image, 
coordinates in the mirror image are related to the original coordinates as 
\begin{align}
 \bar v &= U + v_0 - r_h(v_0) + \mathcal O(G_N) \ , 
 \\ 
 \bar U &= v - v_0 + r_h(v_0) + \mathcal O(G_N) \ . 
\end{align}
Then, $v'_b$ and $U'_b$ are written as 
\begin{align}
 v'_b &= U_b + v_0 - r_h(v_0) + \mathcal O(G_N) \ , 
 \\ 
 U'_b &= v_b - v_0 + r_h(v_0) + \mathcal O(G_N) \ . 
\end{align}
Thus, the entanglement entropy is given by 
\begin{equation}
 S(R) 
 = 
 \frac{\pi r^2(v_b,U_b)}{G_N}
 + \frac{c}{6} \log\left|v_b - v_0 - U_b + r_h(v_0)\right| \ , 
 \label{S-flat}
\end{equation}
for $v<v_0$ and 
\begin{align}
 S(R) 
 = 
 \frac{\pi r^2(v_b,U_b)}{G_N}
 + \frac{c}{6} \log\left|v_b - v_0 - U_b + r_h(v_0)\right| 
 + \frac{c}{24} \int_{v_0}^{v_b} \frac{dv}{r_h(v)} 
 \ , 
 \label{S-before}
\end{align}
near the horizon for $v>v_0$. 

Next, we consider the configuration with the island. 
Since correlations between twist operators on $a$ and $b$ 
dominate over the correlations with their mirror images, 
we can just ignore the mirror image assuming that 
the origin $r=0$ is sufficiently far away from the twist operators. 
Thus, the entanglement entropy is expressed as 
\begin{align}
 S(R) 
&= 
 \frac{\pi r^2(v_a,U_a)}{G_N}
 + \frac{\pi r^2(v_b,U_b)}{G_N}
 + \frac{c}{6} \log\left|\left(v_a - v_b\right)\left(U_a - U_b\right)\right|
\notag\\
&\quad
 + \frac{c}{24} \int_{v_0}^{v_a} \frac{dv}{r_h(v)} 
 + \frac{c}{24} \int_{v_0}^{v_b} \frac{dv}{r_h(v)} 
 + \mathcal O(G_N) \ . 
 \label{S}
\end{align}
The position of the quantum extremal surface $(v_a,U_a)$ is determined 
so that the entanglement entropy $S(R)$ is extremized, namely, 
\begin{align}
 0 
 &= 
 \frac{\partial S}{\partial v_a} 
 = 
 - \frac{\pi r_h^2(v_0)}{2 G_N r_h^2(v_a)} U_a \exp\left[\int_{v_0}^{v_a} \frac{dv}{2r_h(v)}\right] 
 + \frac{c}{48 r_h(v_a)} + \frac{c}{6\left(v_a-v_b\right)} 
 + \mathcal O(G_N) \ , 
 \label{cond-v}
\\
 0 
 &= 
 \frac{\partial S}{\partial U_a} 
 = 
 - \frac{\pi r_h^2(v_0)}{G_N r_h(v_a)} \exp\left[\int_{v_0}^{v_a} \frac{dv}{2r_h(v)}\right] 
 + \frac{c}{6\left(U_a-U_b\right)}
 + \mathcal O(G_N) \ ,  
 \label{cond-U}
\end{align}
where we used \eqref{drh}. 

Since the change of $r_h(v)$ is very slow as \eqref{drh}, 
we assume that 
\begin{equation}
 r_h(v_a) \simeq r_h(v_b) \simeq r_h \ , 
 \label{ansatz1}
\end{equation}
and then, we obtain 
\begin{equation}
 \frac{r_h^2(v_b)}{r_h^2(v_a)} 
 \exp\left[\int_{v_b}^{v_a} \frac{dv}{2r_h(v)}\right] 
 \simeq 
 e^\frac{v_a-v_b}{2r_h} \ . 
 \label{ansatz2}
\end{equation}
Then, the conditions of the quantum extremal surfaces \eqref{cond-v} and \eqref{cond-U} 
can be rewritten as 
\begin{align}
 0 
 &= 
 - \pi U_b V_b 
 + 2 c\,G_N e^{-\frac{v_a-v_b}{2r_h}}
 \left(-\frac{1}{16r_h} + \frac{1}{6(v_a-v_b)}\right) \ , 
 \label{cond-v2}
\\
 0 
 &= 
 \frac{U_a}{U_b} 
 - \frac{8 r_h + (v_a-v_b)}{8 r_h - 3(v_a - v_b)} \ , 
 \label{cond-U2}
\end{align}
where 
\begin{equation}
 V_b = \frac{r_h^2(v_0)}{r_h^2(v_b)} \exp\left[\int_{v_0}^{v_b} \frac{dv}{2r_h(v)}\right] \ . 
\end{equation}
We focus on the regime 
\begin{equation}
 \frac{G_N}{r_h(v_b)} 
 \ll 
 \left|U_b V_b\right|
 \ll r_h(v_b) \ , 
 \label{regime}
\end{equation}
and then, eq.~\eqref{cond-v2} has two solutions 
\begin{align}
 v_a 
 &\simeq 
 v_b - 2 r_h \log\left|\frac{8\pi r_h U_b V_b}{c\,G_N}\right| \ , 
 \label{sol-v}
\\
 v_a 
 &\simeq 
 v_b + \frac{c\,G_N}{3\pi U_b V_b} \ . 
 \label{fake-v}
\end{align}
By substituting these solutions, \eqref{cond-U2} gives 
\begin{align}
 U_a 
 &\simeq 
 - \frac{1}{3} U_b 
 \left[
 1 - \frac{16}{3}\left(\log\left|\frac{8\pi r_h U_b V_b}{c\,G_N}\right|\right)^{-1}
 \right] 
& 
 &\text{for}
&
 v_a 
 &\simeq 
 v_b - 2 r_h \log\left|\frac{8\pi U_b V_b}{c G_N}\right| \ , 
 \label{sol-U}
\\
 U_a 
 &\simeq 
 U_b 
 \left(1 + \frac{c\,G_N}{6\pi r_h U_b V_b}\right) 
& 
 &\text{for}
&
 v_a 
 &\simeq 
 v_b + \frac{c\,G_N}{3\pi U_b V_b} \ . 
 \label{fake-U}
\end{align}
Here, the solution \eqref{sol-v} has larger $|v_a - v_b|$, which is of order of 
\begin{equation}
 \left|v_a-v_b\right| \lesssim r_h \log \frac{r_h^2}{G_N} \ . 
\end{equation}
Then, the difference of $r_h(v)$ is negligible as the change of $r_h(v)$ is very slow as \eqref{drh}. 
Thus, the solutions above are consistent with \eqref{ansatz1}. 

Now, we have two solutions of \eqref{cond-v} and \eqref{cond-U} --- 
one is given by \eqref{sol-v} and \eqref{sol-U}, and the other is given by \eqref{fake-v} and \eqref{fake-U}. 
The entanglement entropy is given by the smaller value of these two solutions. 
For $|U_b V_b|\sim \mathcal O(G_N^0)$, 
\eqref{sol-v} and \eqref{sol-U} give 
\begin{equation}
 r(v_a,U_a) - r_h(v_a) - 2 r_h(v_a) \dot r_h(v_a) 
 =
 - \frac{r_h^2(v_0)}{2r_h^2(v_a)} U_a \exp\left[\int_{v_0}^{v_a} \frac{dv}{2r_h(v)}\right] 
 \simeq - \frac{c\,G_N}{48\pi r_h(v_a)} \ , 
\end{equation}
and hence, the quantum extremal surface $a$ is located inside the event horizon 
and the distance from the event horizon is within the order of the Planck length. 
By using \eqref{drh}, the area of the apparent horizon obeys 
\begin{equation}
 4 \pi r_h^2(v) 
 = 
 4 \pi r_h^2(v_0) - c\,G_N \int_{v_0}^{v} \frac{dv'}{12 r_h(v')} \ .  
\end{equation}
Then, the area terms in \eqref{S} are expressed as 
\begin{align}
 \frac{\pi r^2(v_a,U_a)}{G_N}
 &= 
 \frac{\pi r_h^2(v_0)}{G_N} 
 + \frac{c}{24} \log\left|\frac{8\pi r_h^2(v_0) U_b}{c\,G_N r_h(v_a)}\right| 
 - \frac{c}{12}
 \ . 
 \label{area-a}
\\
 \frac{\pi r^2(v_b,U_b)}{G_N}
 &= 
 \frac{\pi r_h^2(v_0)}{G_N} 
 - \frac{\pi r_h(v_b) U_b V_b}{G_N}
 - \int_{v_0}^{v_b} \frac{c\,dv}{48 r_h(v)} 
 - \frac{c}{24}
 \ . 
 \label{area-b}
\end{align}
The non-local terms are estimated as 
\begin{align}
 S_\text{matter}^\text{(non-local)}
&\simeq 
 \frac{c}{6} \log\left|\left(v_a - v_b\right)\left(U_a - U_b\right)\right|
 + \frac{c}{24} \int_{v_0}^{v_a} \frac{dv}{r_h(v)} 
 + \frac{c}{24} \int_{v_0}^{v_b} \frac{dv}{r_h(v)} 
\\
&\simeq
 \frac{c}{6} \log\left[\frac{8 r_h}{3}\left|U_b\right|\log\left|\frac{8\pi r_h U_b}{c\,G_N}\right|\right]
 - \frac{c}{12} \log\left|\frac{8\pi r_h U_b}{c\,G_N}\right| 
 + \frac{c}{24} \int_{v_0}^{v_b} \frac{dv}{r_h(v)} \ . 
\end{align}
Thus, the entanglement entropy is obtained as 
\begin{equation}
 S(R) 
 \simeq 
 \frac{2 \pi r_h^2(v_0)}{G_N} - \frac{\pi r_h(v_b) U_b V_b}{G_N} 
 + \frac{c}{8}\log\left|U_b\right| + \int_{v_0}^{v_b} \frac{c\,dv}{48 r_h(v)} \ , 
 \label{S-far}
\end{equation}
where we have ignored constant terms of $\mathcal O(G_N^0)$ and all terms of $\mathcal O(G_N)$. 

For the other solution \eqref{fake-v} and \eqref{fake-U}, the quantum extremal surface $a$ is 
located within a distance of $\mathcal O(G_N)$ from the (inner) boundary $b$ of the region $R$. 
\begin{equation}
 S(R) 
 \simeq 
 \frac{2 \pi r_h^2(v_0)}{G_N} 
 - \frac{2\pi r_h(v_b) U_b V_b}{G_N}
 - \int_{v_0}^{v_b} \frac{c\,dv}{24 r_h(v)} \ , 
 \label{S-fake}
\end{equation}
where constant terms of $\mathcal O(G_N^0)$ are ignored, again. 
Near the apparent horizon $|U_b V_b|\ll r_h$, we have 
\begin{equation}
 \log\left|U_b\right| \ll - \int_{v_0}^{v_b} \frac{dv}{2r_h(v)} \ . 
\end{equation}
Since $U_b V_b < 0$, the solution \eqref{S-far} 
is smaller than \eqref{S-fake}, and hence, 
$a$ is located at \eqref{sol-v} and \eqref{sol-U}. 
The entanglement entropy is given by \eqref{S-far} 
for $|U_b V_b|\sim\mathcal O(G_N^0)$ and $|U_b V_b|\ll r_h(v_b)$. 

For smaller $|U_b V_b|$, two solutions above approaches to each other. 
Two solutions \eqref{sol-v} and \eqref{fake-v} become the same to each other, 
or equivalently, \eqref{cond-v2} has a double root, for 
\begin{equation}
 U_b V_b = - \frac{3c\,G_N}{8\pi r_h} e^{2/3} \ , 
 \label{crit-b}
\end{equation}
and the solution is given by 
\begin{align}
 v_a &= v_b - \frac{4}{3} r_h \ , 
&
 U_a = \frac{5}{9} U_b \ . 
 \label{crit-a}
\end{align}
The entanglement entropy is given by 
\begin{equation}
 S(R) 
 \simeq 
 \frac{2 \pi r_h^2(v_b)}{G_N} 
 \simeq 
 \frac{2 \pi r_h^2(v_0)}{G_N} - \int_{v_0}^{v_b} \frac{c\,dv}{24 r_h(v)} \ . 
 \label{S-crit}
\end{equation}
For smaller $|U_b V_b|$, there is no saddle point for the quantum extremal surface, 
since the solution of $v_a$ becomes complex. 
At the critical point \eqref{crit-b}, the separation between 
the island $I$ and the region $R$ is of the order of the Planck length, 
and hence, it would be reasonable that the entanglement entropy cannot be calculated 
by using the island rule, as it is valid only in the semi-classical regime. 
A possible interpretation is that the entanglement entropy becomes zero beyond this critical point. 
Since the UV cutoff of the semi-classical gravity should be around the Planck scale, 
two points $a$ and $b$ is almost indistinguishable in the semi-classical regime. 
For smaller $|U_b V_b|$, the region $R$ and island $I$ are effectively merged 
and becomes a single connected region, and then, 
\begin{equation}
 S(R) = 0 \ . 
\end{equation}


\subsection{Page time as an outgoing null surface}\label{ssec:Page}

In this paper, we calculate the quantum expansion \eqref{QE} along outgoing null lines, 
and spherically symmetric cross-sections of the null line is identified 
with the inner boundary $b$, and then, 
the generalized entropy is given by the entanglement entropy, 
\begin{equation}
 S_\text{gen}=S(R) \ . 
 \label{Sgen=SR}
\end{equation}
For this purpose, we consider time evolution of $b$ along the outgoing null surface, $U = \text{const}$. 
Before that, we first see how the conventional Page curve is reproduced by the island rule. 

The Page curve is reproduced by considering time evolution 
along a timelike surface near the apparent horizon. 
We consider the surface which satisfies $U_b V_b = \text{const}$. 
The surface is parameterized by $v_b$, and $U_b$ is expressed in terms of $v_b$ as 
\begin{equation}
 \log\left|U_b\right| = \log\left|U_0\right| - \int_{v_0}^{v_b}\frac{dv}{2r_h(v)} \ , 
\end{equation}
where $U_0$ is the retarded time at the intersection of the timelike surface 
$U_b V_b = U_0 = \text{const.}$ and the collapsing null shell on $v=v_0$. 
The entanglement entropy for the configuration with the island becomes 
\begin{equation}
 S(R) 
 \simeq 
 \frac{2 \pi r_h^2(v_0)}{G_N} - \frac{\pi r_h(v_b) U_0}{G_N} 
 - \int_{v_0}^{v_b} \frac{c\,dv}{24 r_h(v)} \ . 
 \label{S-temp}
\end{equation}
Thus, the entanglement entropy decreases with time. 

\begin{figure}[t]
\begin{center}
\includegraphics[scale=0.7]{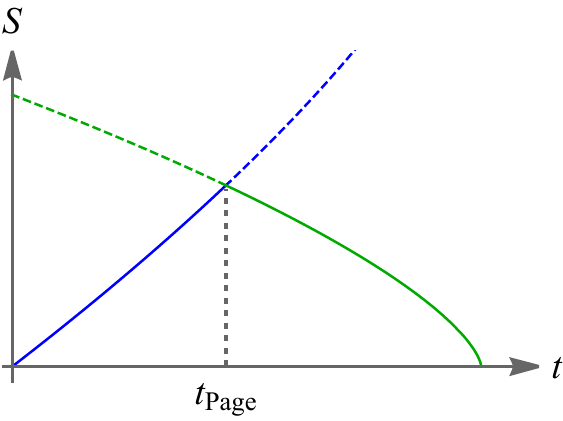}
\hspace{24pt}
\includegraphics[scale=0.7]{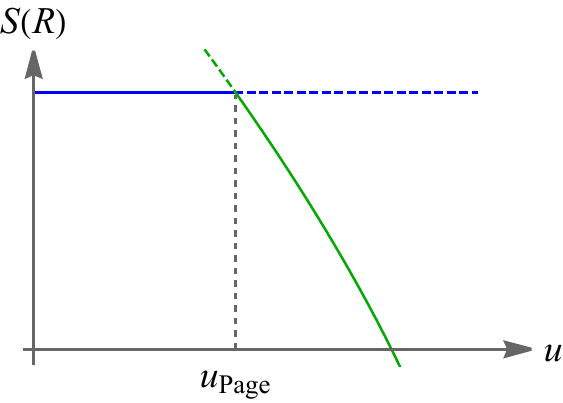}
\caption{%
The Page curve along a timelike surface $U_b V_b=\text{const.}$ (left) and 
an ingoing null surface $v=\text{const.}$ (right). 
The conventional Page curve is reproduced by considering 
the entanglement entropy of the region $R$, whose boundary $b$ moves along a timelike surface. 
Since \eqref{S(R)} has additional area term of $b$, 
the Page curve is obtained by subtracting it; $S=S(R) - \frac{\pi r^2(v_b,U_b)}{G_N}$. 
If $b$ moves along an ingoing null surface $v=\text{const.}$, 
the entanglement entropy no longer gives the conventional Page curve, 
but the Page time can be read off from the curve. 
If $b$ moves along an outgoing null surface $U=\text{const.}$, 
there is no Page time on the surface, and the entanglement entropy 
always increases as $v$ increases. 
}\label{fig:curve}
\end{center}
\end{figure}

For the configuration without islands, 
by using \eqref{area-b}, the entanglement entropy \eqref{S-before} becomes 
\begin{equation}
 S(R) 
 \simeq 
 \frac{\pi r_h^2(v_0)}{G_N} - \frac{\pi r_h(v_b) U_b V_b}{G_N} 
 + \frac{c}{6} \log\left|v_b - v_0 - U_b + r_h(v_0)\right| 
 + \int_{v_0}^{v_b} \frac{c\,dv}{48 r_h(v)} \ . 
 \label{S-before2}
\end{equation}
The entanglement entropy increases with time in this configuration 
and becomes larger than \eqref{S-temp} if 
\begin{equation}
 \int_{v_0}^{v_b} \frac{c\,dv}{16 r_h(v)} > \frac{\pi r_h^2(v_0)}{G_N} \ , 
 \label{Page-V}
\end{equation}
where we ignored $\mathcal O(G_N^0)$ terms. 
Thus, \eqref{S-before} is smaller than \eqref{S-temp} for $v-v_0<t_\text{Page}$, 
and \eqref{S-temp} is smaller than \eqref{S-before} for $v-v_0>t_\text{Page}$, 
where the Page time $t_\text{Page}$ is given by 
\begin{equation}
 t_\text{Page} \simeq \frac{32\pi r_h^3(v_0)}{c\,G_N}\left(1-\frac{2^{3/2}}{3^{3/2}}\right) 
 \approx \frac{15\pi r_h^3(v_0)}{c\,G_N} \ . 
 \label{t-Page}
\end{equation}
The entanglement entropy is given by \eqref{S-flat} for $v_b<v_0$, 
\eqref{S-before2} for $v_0 < v_b < v_0 + t_\text{Page}$ 
and \eqref{S-temp} for $v_b > v_0 + t_\text{Page}$. 
Thus, the Page curve is reproduced by the island rule (Fig.~\ref{fig:curve} (left)). 

Now, we consider the time evolution of $b$ along the outgoing null surface. 
The transition from the configuration without islands to the configuration with the island 
occurs when the entanglement entropy \eqref{S-before2} exceeds \eqref{S-far}, namely, 
\begin{equation}
 \frac{c}{6} \log\left|v_b - v_0 - U_b + r_h(v_0)\right| 
 > \frac{\pi r_h^2(v_0)}{G_N} + \frac{c}{8}\log\left|U_b\right| \ . 
 \label{cond-Page}
\end{equation}
Since $U_b=\text{const.}$ on an outgoing null line, 
the r.h.s.\ of \eqref{cond-Page} is also a constant, 
and hence, \eqref{cond-Page} will be satisfied for sufficiently large $v$. 
However, if $\log|U_b|\sim\mathcal O(G_N)$, 
\eqref{cond-Page} is satisfied only for $v\sim \exp\left[\mathcal O(G_N^{-1})\right]$. 
As the black hole will evaporate in the lifetime of 
\begin{equation}
 t_\text{eva} \simeq \frac{32\pi r_h^3(v_0)}{c\,G_N} \ , 
 \label{t-eva}
\end{equation}
the condition \eqref{cond-Page} cannot be satisfied during the lifetime of the black hole. 
Without the fine-tuning of $U_b$, no transition from \eqref{S-before2} to \eqref{S-far} 
occurs along the outgoing null surface. 
Along outgoing null surfaces, or equivalently for $U_b=\text{const.}$, 
the entanglement entropy \eqref{S-before2} is always increasing with time. 

Since the lifetime of the black hole \eqref{t-eva} 
is $t_\text{eva} \sim \mathcal O(G_N^{-1})$, 
the l.h.s.\ of \eqref{cond-Page} is negligible 
compared with the area term in the r.h.s. 
Thus, the condition of the Page time is simply expressed as 
\begin{equation}
 -\frac{c}{8}\log\left|U_b\right| > \frac{\pi r_h^2(v_0)}{G_N} \ . 
 \label{Page-U}
\end{equation}
Although no transition occurs along outgoing null surfaces, 
the configuration with the island is realized on null surfaces with sufficiently small $|U_b|$. 
The entanglement entropy \eqref{S-far} is increasing with time 
along outgoing null surfaces $U_b=\text{const}$. 
Thus, each outgoing null surface lies either before or after the Page time, 
but cannot extend across the Page time, 
and the entanglement entropy is increasing with time in either sides of the Page time. 

\begin{figure}[t]
\begin{center}
\includegraphics[scale=0.4]{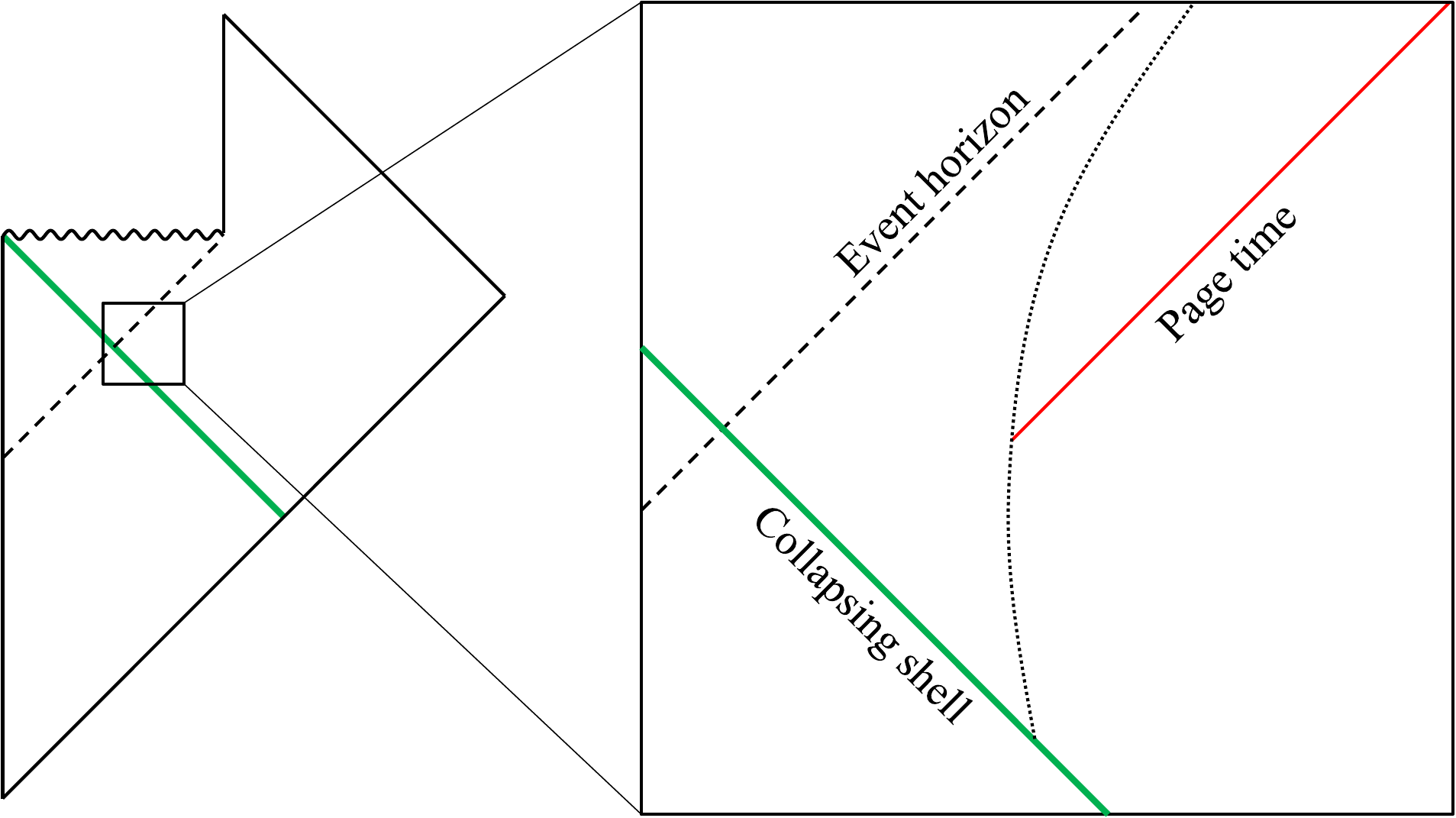}
\caption{%
The Page time as a null surface. 
We define the Page time by the position of the surface $b$ where 
the dominant saddle changes from the configuration without islands 
to the configuration with the island. 
Then, the Page time is expressed as an approximately null surface. 
The entanglement entropy cannot be calculated by using the island rule 
if the surface $b$ is too close to the event horizon, 
which is indicated by the region inside the dotted line. 
}\label{fig:Page}
\end{center}
\end{figure}

The Page time is expressed as a line in the two-dimensional space of $(v_b,U_b)$, 
or equivalently, three dimensional surface in the ingoing Vaidya spacetime. 
To be exact, it is a spacelike timeslice but approximately a null surface 
at the leading order of the small-$G_N$ expansion. 
Thus, it is more suitable to specify the Page time in the retarded time $U_b$, 
which is nothing but \eqref{Page-U}. 
In the case of the timelike surface with $|U_b V_b|\sim\mathcal O(G_N^0)$, 
the Page time is expressed in terms of $v_b$ by virtue of the condition $U_b V_b = \text{const.}$; 
\begin{equation}
 -\frac{c}{8}\log\left|U_b\right| 
 = 
 - \frac{c}{8}\log\left|U_b V_b\right| 
 + \int_{v_0}^{v_b} \frac{c\,dv}{16 r_h(v)} \ , 
\end{equation}
and then, \eqref{cond-Page} reproduces \eqref{Page-V} 
as $\log\left|U_b V_b\right|\sim\mathcal O(G_N^0)$ is negligible compared with the Page time. 
The surface of the Page time is very close to the event horizon for $v\simeq v_0$ 
and enters the region \eqref{regime} for $v_b \simeq v_0 + t_\text{Page}$. 
Thus, the Page time is approximately the same for any timelike surface in this region. 

Although, the Page time is expressed as \eqref{Page-U}, 
it would be more convenient to use the retarded time of an observer in the future null infinity. 
Near the future null infinity, the energy-momentum tensor approximately given by 
\begin{align}
 T_{uu} 
 &\simeq 
 \frac{c}{768\pi^2 r_h^2 r^2} \ , 
 &
 T_{vv} 
 &\simeq 
 0 \ . 
\end{align}
Thus, the spacetime is approximated by the outgoing Vaidya spacetime, 
\begin{equation}
 ds^2 = - \left(1 - \frac{\bar r_h(u)}{r}\right) du^2 - 2du\,dr + r^2 d \Omega^2 \ , 
\end{equation}
where $\bar r_h(u)$ is determined by 
\begin{align}
 \bar r_h(u_0) 
 &= 
 r_h(v_0) \ , 
 &
 \frac{d\bar r_h(u)}{du} 
 &= 
 - 8\pi G_N r^2 T_{uu} \simeq - \frac{c\,G_N}{96\pi \bar r_h(u)^2} \ , 
\end{align}
Then, the retarded time near the apparent horizon $U$ 
would be related to retarded time near the future null infinity $u$ as%
\footnote{%
The relation can roughly be read off from the classical Schwarzschild solution. 
For more detailed relation in the evaporating black hole geometry, 
the metric near the horizon can be expressed by using the outgoing Vaidya metric 
with higher order corrections of the semi-classical approximation \cite{Ho:2018jkm,Ho:2019pjr}.
} 
\begin{equation}
 U \simeq - \bar r_h(u) \exp\left[-\int_{u_0}^u \frac{du'}{2\bar r_h(u)}\right] \ . 
\end{equation}
Then, the condition \eqref{Page-U} is rewritten as 
\begin{equation}
 \int_{u_0}^{u_b} \frac{c\,du}{16 \bar r_h(u)} > \frac{\pi r_h^2(v_0)}{G_N} \ . 
\end{equation}
The Page time in the retarded time $u$ is given by the same expression to \eqref{t-Page},%
\footnote{%
Although it would be more appropriate to define the Page time in the retarded time, 
the entanglement entropy does not give the conventional form of the Page curve 
when $b$ moves along ingoing null surfaces. 
} 
\begin{equation}
 u_\text{Page} \simeq t_\text{Page} \simeq \frac{32\pi r_h^3(v_0)}{c\,G_N}\left(1-\frac{2^{3/2}}{3^{3/2}}\right) 
 \approx \frac{15\pi r_h^3(v_0)}{c\,G_N} \ .  
 \label{u-Page}
\end{equation}
To be more precise, the argument above is 
valid only for $|U|\ll r_h(v_0)$, or equivalently $u\gg u_0$. 
Since the collapsing shell reaches this region within the scrambling time, 
\begin{equation}
 t_\text{scr} \simeq 2r_h \log \frac{r_h^2}{G_N} \ , 
\end{equation}
contributions from $u\lesssim u_0$ are negligible in the calculation of the Page time above. 
In the argument above, we have ignored higher order corrections, 
which do not affect the result for the Page time. 

So far, we have considered only the cases of $|U_b V_b|\ll r_h(v_0)$. 
Outgoing null lines leave this region, for sufficiently large $v_b$. 
If $|U_b V_b|\ll r_h(v_0)$ at some point on an outgoing null surface after the collapse of the shell, 
the retarded time satisfies $|U_b|\ll r_h(v_0)$ on the surface. 
Then, the conditions of the quantum extremal surface, \eqref{cond-v} and \eqref{cond-U}, 
which determine the position of $a$, are not modified even for $|U_b V_b| \gtrsim r_h(v_0)$. 
Thus, the entanglement entropy is given by 
\begin{equation}
 S(R) 
 = 
 \frac{\pi r_h^2(v_b)}{G_N} 
 + \frac{c}{6} \log\left|v_b - v_0 - U_b + r_h(v_0)\right| 
 + \frac{c}{12}\log\left|\frac{\partial r(v_b,U_b)}{\partial U_b}\right| \ , 
 \label{S-away0}
\end{equation}
for the configuration without islands, and 
\begin{equation}
 S(R) 
 \simeq 
 \frac{\pi r_h^2(v_0)}{G_N} + \frac{\pi r^2(v_b,U_b)}{G_N} 
 + \frac{c}{8}\log\left|U_b\right| 
 + \frac{c}{12}\log\left|\frac{\partial r(v_b,U_b)}{\partial U_b}\right| \ , 
 \label{S-away1}
\end{equation}
for the configuration with the island. 
Here, the radius $r(v,U)$ is given by \eqref{r-full}. 
Since $r(v,U)$ and $\frac{\partial r(v,U)}{\partial U}$ is 
monotonically increasing for $|U_b V_b|\gtrsim r_h(v_0)$, 
the entanglement entropy is always increasing with time along 
outgoing null surfaces with $|U_b|\ll r_h(v_0)$. 
For $V_b \gg 1$, $\frac{\partial r(v,U)}{\partial U}$ approaches 
\begin{equation}
 \frac{\partial r(v,U)}{\partial U} \to \frac{r_h(v_b)}{U_b} \ . 
\end{equation}
Thus, the matter part of the entanglement entropy is 
increasing near the apparent horizon $|U_b V_b|\ll r_h(v_b)$ as 
\begin{equation}
 S_\text{matter}^\text{(non-local)} 
 \simeq \frac{c}{12}\log|U_b| + \int_{v_0}^{v_b} \frac{c\,dv}{16 r_h(v)} \ , 
\end{equation}
but is bounded from above by 
\begin{equation}
 S_\text{matter}^\text{(non-local)} 
 \simeq 
 \frac{c}{12}\log\left|U_b\right| 
 + \frac{c}{12}\log\left|\frac{\partial r(v_b,U_b)}{\partial U_b}\right| < 0 \ . 
\end{equation}
Note that the matter part of the entanglement entropy is negative because 
we have ignored constant terms. 
In particular, the UV cutoff is defined so that 
the entanglement entropy becomes zero when 
the separation between two endpoints becomes as small as the cutoff scale. 
Thus, the entanglement entropy is positive by definition if 
constant terms are taken into account.


\subsection{Quantum focusing conjecture for evaporating black holes}\label{ssec:QFC}

Now, we study the quantum focusing conjecture 
along outgoing null surfaces around the evaporating black hole. 
As we have seen in Sec.~\ref{ssec:violation}, 
the violation of the classical focusing condition 
can easily be seen from the fact that 
the expansion on an outgoing null line becomes negative 
although the null line reaches the future null infinity. 

For the quantum focusing conjecture, 
the quantum expansion is defined by using the entanglement entropy, 
namely \eqref{QE} with \eqref{Sgen=SR}. 
In Sec.~\ref{ssec:Page}, we have considered the time evolution of the entanglement entropy, 
and found that the entanglement entropy is always increasing with time 
along outgoing null surfaces for $|U_b V_b|\gg G_N/r_h$. 
For the configuration without islands, the expression \eqref{S-before2} 
is valid also for smaller $|U_b V_b|$, and hence, the quantum expansion is always positive. 

For the configuration with the island, it is easier to calculate 
the quantum expansion directly from \eqref{S}. 
The quantum expansion is expressed as 
\begin{equation}
 \Theta = \frac{1}{4\pi r^2} \frac{dS(R)}{d \lambda} 
 = 
 \frac{1}{4\pi r^2} \frac{dv_b}{d \lambda} 
 \left(
 \frac{\partial S(R)}{\partial v_b} 
 + \frac{dv_a}{dv_b} \frac{\partial S(R)}{\partial v_a} 
 + \frac{dU_a}{dv_b} \frac{\partial S(R)}{\partial U_a} 
 \right) \ . 
\end{equation}
Since the quantum extremal surface satisfies 
\begin{equation}
 \frac{\partial S(R)}{\partial v_a} = \frac{\partial S(R)}{\partial U_a} = 0 \ ,  
 \label{cond-QES}
\end{equation}
the quantum expansion is positive if $\frac{\partial S(R)}{\partial v_b}$ is positive. 
By using \eqref{S} with \eqref{drh} and \eqref{r}, we obtain 
\begin{equation}
 \frac{\partial S(R)}{\partial v_b} 
 \simeq 
 \frac{c}{48 r_h(v_b)} 
 - \frac{\pi r_h^2(v_0)}{2r_h^2(v_b)} U_b V_b + \frac{c}{6\left(v_b-v_a\right)} > 0 \ . 
\end{equation}
Thus, the entanglement entropy is always increasing with time 
and the quantum expansion is positive. 

Now, we consider the quantum focusing condition \eqref{QFC}. 
For the configuration without islands, the derivative of the quantum expansion is calculated as 
\begin{align}
 \frac{d\Theta}{d \lambda} 
 &\simeq 
 - \frac{U_b^2}{8 G_N r_h^4(v_b)} 
 - \frac{c}{96\pi r_h^4(v_0)V_b^2} 
\notag\\
&\quad
 - \frac{c\,r_h^2(v_b)}{6\pi r_h^4(v_0)(v_b-v_0 - U_b + r_h(v_0))^2V_b^2} 
\notag\\
&\quad
 - \frac{c\,r_h(v_b)}{12\pi r_h^3(v_0)(v_b-v_0 - U_b + r_h(v_0))V_b^2} 
 \ . 
\end{align}
Since the outgoing null line intersects the origin $r=0$ at 
$v = v_0 + U_b - r_h(v_0)$, the quantum expansion is satisfied 
for the configuration without islands, 
\begin{equation}
 \frac{d\Theta}{d \lambda} < 0 \ . 
\end{equation}

For the configuration with the island, 
the derivative of the quantum expansion is rewritten as 
\begin{align}
 \frac{d\Theta}{d \lambda} 
 &= 
 \frac{d}{d \lambda} \left(\frac{1}{4\pi r^2} \frac{dS(R)}{d \lambda}\right) 
 = 
 \frac{d}{d \lambda} 
 \left(\frac{1}{4\pi r^2} \frac{dv_b}{d \lambda} \frac{\partial S(R)}{\partial v_b} \right)
 \ , 
\end{align}
where we have used \eqref{cond-QES} and its derivative, namely, 
\begin{equation}
 \frac{d}{d \lambda} \frac{\partial S(R)}{\partial v_a}
 = 
 \frac{d}{d \lambda} \frac{\partial S(R)}{\partial U_a}
 = 0 \ , 
\end{equation}
which follows from the condition that \eqref{cond-QES} is satisfied 
everywhere on the outgoing null surface. 
Then, the quantum expansion is calculated as 
\begin{align}
 \frac{d\Theta}{d \lambda} 
 &\simeq 
 - \frac{U_b^2}{8 G_N r_h^4(v_b)} 
 - \frac{c}{96\pi r_h^4(v_0)V_b^2} 
\notag\\
&\quad
 - \frac{c\,r_h^2(v_b)}{6\pi r_h^4(v_0)(v_b-v_a)^2V_b^2} 
 - \frac{c\,r_h(v_b)}{12\pi r_h^3(v_0)(v_b-v_a)V_b^2} 
\notag\\
&\quad
 + \frac{c\,r_h^2(v_b)}{6\pi r_h^4(v_0)(v_b-v_a)^2V_b^2}\frac{dv_a}{dv_b} \ . 
\end{align}
From \eqref{cond-v}, we obtain 
\begin{equation}
 \frac{dv_a}{dv_b} = \frac{16r_h}{\left[4r_h+3(v_a-v_b)\right]\left[4r_h-(v_a-v_b)\right]} \ . 
\end{equation}
The conditions \eqref{cond-v} and \eqref{cond-U} have two solutions at 
$v_a - v_b < - \frac{3}{4}r_h$ and $v_a - v_b > - \frac{3}{4}r_h$, 
but the solution with $v_a - v_b < - \frac{3}{4}r_h$ gives 
a smaller value of the entanglement entropy as we have seen in Sec.~\ref{ssec:island}. 
Therefore, we have 
\begin{equation}
 \frac{dv_a}{dv_b} < 0 \ , 
\end{equation}
and hence, the quantum focusing conjecture is satisfied, 
\begin{equation}
 \frac{d\Theta}{d \lambda} < 0 \ .  
\end{equation}


\subsection{Beyond the s-wave approximation}\label{ssec:momenta}

In this paper, we have focused only on the s-wave 
and ignored the other modes with non-zero angular momentum. 
Here, we make a rough estimation of corrections from these modes. 

Since the geometry has the spherical symmetry, 
massless free fields in four-dimensional spacetime can be expanded 
by using the spherical harmonics, or equivalently, 
into eigenmodes of the angular momentum. 
Each modes can be treated as independent fields without interactions. 
The twist operator, which is introduced for the replica trick, 
is given by the product of the twist operators of each independent mode, 
and the entanglement entropy becomes the sum of contributions of each mode, 
\begin{equation}
 S = \sum_{l,m} S_{lm} \ , 
\end{equation}
where $S_{lm}$ is the entanglement entropy of the mode with the angular momentum $(l,m)$ 
and formally expressed (in a similar fashion to \eqref{Smatter0}) as 
\begin{equation}
  S_{lm}  
 = 
 \frac{c}{6} \sum_{i,j}D_{lm}(x_i,x'_j) 
 - \frac{c}{12} \sum_{i,j}D_{lm}(x_i,x_j) 
 - \frac{c}{12} \sum_{i,j}D_{lm}(x'_i,x'_j) \ , 
\end{equation}
where $D_{lm}(x,y)$ comes from the correlation function of the twist operators.%
\footnote{%
Note that $D_{lm}(x,y)$ is different from the correlation function of the field itself, in general. 
} 
The modes with angular momenta can be treated as Kaluza-Klein modes 
in two-dimensional spacetime of temporal and radial directions, 
and the Kaluza-Klein mass of each mode becomes negligible when 
we consider the correlation in very small distance. 
Thus, the two-point function $D_{lm}(x,y)$ should approach 
the same value to the case of the s-wave, 
\begin{align}
 D_{lm}(x,y) &\to \log|x-y|^2 \ , 
\end{align}
when the two points are very close to each other, $y\to x$. 
Thus, the modes with angular momenta give the same term to the s-wave from the UV cutoff, 
and the entanglement entropy of these modes is expressed in a similar fashion to \eqref{Smatter} as 
\begin{align}
 S_{lm}  
 &= 
 \frac{c}{6} \sum_{i,j}D_{lm}(x_i,x'_j) 
 - \frac{c}{6} \sum_{i<j}D_{lm}(x_i,x_j) 
 - \frac{c}{6} \sum_{i<j}D_{lm}(x'_i,x'_j) \ , 
\notag\\&\quad 
 + \frac{c}{12} \sum_i \left|\frac{4g_{uv}(x_i)g_{uv}(x'_i)}{\epsilon^4}\right| \ . 
\end{align}

Thus, the entanglement entropy is given by 
\begin{align}
 S(R) 
&= 
 \frac{\pi r^2(v_a,U_a)}{G_N}
 + \frac{\pi r^2(v_b,U_b)}{G_N}
 + \frac{c}{6} \sum_{l,m} D_{lm}(x_a,x_b) 
\notag\\
&\quad
 + \frac{c_\text{EE}}{24} \int_{v_0}^{v_a} \frac{dv}{r_h(v)} 
 + \frac{c_\text{EE}}{24} \int_{v_0}^{v_b} \frac{dv}{r_h(v)} 
 \ . 
 \label{S4D}
\end{align}
Here, modes with angular momenta much larger than the scale of 
the distance between surfaces $a$ and $b$ 
will not contribute to the entanglement entropy, 
and $c_\text{EE}$ is the effective central charge, or equivalently, 
the number of modes with non-trivial contributions. 
The change of the horizon radius \eqref{drh} also has 
effects of the additional modes, and 
the central charge becomes larger than the s-wave approximation, 
\begin{equation}
 \dot r_h(v) = - \frac{c_\text{HR}G_N}{96\pi r_h^2(v)} \ . 
\end{equation}
Here, only the modes which can reach the asymptotic infinity 
contribute to $\dot r_h(v)$, and hence, 
the effective central charge is the same or less than that in the entanglement entropy, 
\begin{equation}
 c_\text{HR} < c_\text{EE} \ . 
\end{equation}
By using \eqref{S4D}, the derivative of the quantum expansion is estimated as 
\begin{align}
 \frac{d\Theta}{d \lambda} 
 &\simeq 
 - \frac{U_b^2}{8 G_N r_h^4(v_b)} 
 - \frac{2c_\text{EE}-c_\text{HR}}{192\pi r_h^4(v_0)V_b^2} 
\notag\\
&\quad
 + \frac{c_\text{EE}\,r_h^2(v_b)}{6\pi r_h^4(v_0)V_b^2} 
   \sum_{l,m} \frac{\partial^2}{\partial v_b^2} D_{lm}(x_a,x_b) 
 - \frac{c_\text{EE}\,r_h(v_b)}{12\pi r_h^3(v_0)V_b^2} 
   \sum_{l,m} \frac{\partial}{\partial v_b} D_{lm}(x_a,x_b) 
\notag\\
&\quad
 + \frac{c_\text{EE}\,r_h^2(v_b)}{6\pi r_h^4(v_0)V_b^2} \frac{d v_a}{d v_b}
   \sum_{l,m} \frac{\partial^2}{\partial v_a \partial v_b} D_{lm}(x_a,x_b) \ . 
\label{dTheta4D}
\end{align}
Since the correlation function $D_{lm}(x,y)$ has asymptotic behaviors, 
\begin{align}
 D_{lm}(x,y) &\to - \infty \ , & (x&\to y) \ , 
 \label{D-infty}
\\
 D_{lm}(x,y) &\to \text{const.} \ , & (x&\to\infty) \ . 
\end{align}
it is expected to satisfy the following conditions 
\begin{align}
 & \partial_x D_{lm}(x,y) = - \partial_y D_{lm}(x,y) > 0 \ , &&(\text{for}\quad x>y)
\label{dD}
\\
 & \partial_x^2 D_{lm}(x,y) < 0 \ , 
\label{ddD}
\end{align}
assuming that $D_{lm}(x,y)$ has sufficiently simple behaviors. 

We have seen in Sec.~\ref{ssec:QFC}, 
the quantum extremal surface satisfies $\frac{dv_a}{dv_b} < 0$ in the s-wave approximation. 
This condition is true for more general quantum extremal surfaces in black hole geometries. 
The condition \eqref{D-infty} guarantees that 
the entanglement entropy has the negative divergence 
$S(R) \to -\infty$ in $v_a\to v_b$,%
\footnote{%
The negative divergence of the entanglement entropy 
implies that the entanglement entropy becomes zero 
if the two points are in a distance of the cut-off scale. 
Results in smaller distance is not reliable and 
the physical entanglement entropy does not become negative. 
} 
implying that \eqref{S4D} as a function of $v_a$ has 
a local maximum between the local minimum and $v_a\to v_b$. 
The position of $b$ has a critical point where 
the local minimum and local maximum become identical to each other, 
and there is no saddle point of the quantum extremal surface for smaller values of $v_b$. 
Thus, we have $\frac{dv_b}{dv_a} = 0$ at the critical point. 
Since the local minimum of \eqref{S4D} has smaller value of $v_a$ than the local maximum, 
we have $\frac{dv_b}{dv_a} < 0$ for the local minimum and $\frac{dv_b}{dv_a} > 0$ for the local maximum. 
Therefore, the quantum extremal surface, which is a saddle point with smaller $S(R)$, satisfies 
\begin{equation}
 \frac{dv_a}{dv_b} < 0 \ . 
 \label{dvdv4D}
\end{equation}

By using the conditions \eqref{dD}, \eqref{ddD} and \eqref{dvdv4D}, 
the expression \eqref{dTheta4D} shows that the quantum focusing conjecture is satisfied, 
\begin{equation}
 \Theta < 0 \ . 
\end{equation}

If the inner boundary $b$ of the region $R$ is sufficiently close to the horizon, 
or equivalently to the quantum extremal surface $a$, 
the entanglement entropy is simply expressed as \cite{Casini:2005zv,Casini:2009sr} 
\begin{align}
 S(R) 
&= 
 \frac{\pi r^2(v_a,U_a)}{G_N}
 + \frac{\pi r^2(v_b,U_b)}{G_N}
 - \frac{4\pi \kappa c r^2(v_b,U_b)}{L^2(x_a,x_b)} 
 \ , 
 \label{S4Dnear}
\end{align}
where $\kappa$ is some constant, and $L(x_a,x_b)$ is the proper distance between $a$ and $b$. 
As we have discussed in Sec.~\ref{ssec:island}, 
the entanglement entropy of matters \eqref{Smatter} 
does not have the covariant expression and 
depends on coordinates which are used to define the creation and annihilation operators 
(See also Appendix~\ref{app:entropy}). 
However, universal divergences in correlation functions dominates in small distances, 
and then, the entanglement entropy is approximated by covariant terms. 
The entanglement entropy \eqref{S} in the s-wave approximation, 
in fact, can be approximated by the expression in terms of the proper distance, 
\begin{equation}
 L^2 
 \simeq 
 V_b \left|(U_a-U_b)(v_a - v_b)\right|
 \simeq 
 2r_h(v_b) \left|(U_a - U_b)(V_a - V_b)\right| \ , 
\end{equation}
if $a$ and $b$ are sufficiently close to each other to satisfy the following conditions, 
\begin{align}
 \left|v_a-v_b\right| &\ll 2r_h \ , 
 & 
 V_b \left|(U_a-U_b)(v_a - v_b)\right| &\ll r_h^2 \ . 
 \label{VeryShort}
\end{align}
In such small distances, state-dependent parts of the entanglement entropy become negligible 
and the entanglement entropy for four-dimensional massless fields is expressed as \eqref{S4Dnear}. 
Then, the quantum expansion is calculated as 
\begin{align}
 \frac{d\Theta}{d \lambda} 
 &\simeq 
 - \frac{U_b^2}{8 G_N r_h^4(v_b)} 
 + \frac{c_\text{HR}}{96\pi r_h^4(v_0)V_b^2} 
 + \frac{\kappa c\,r_h(v_b)}{r_h^4(v_0)(V_a-V_b)^3(U_a-U_b)} \left(1-\frac{dv_a}{dv_b}\right)
 \ . 
\label{dTheta4Dnear}
\end{align}
Here, \eqref{dvdv4D} is satisfied as we have discussed. 
By using the conditions \eqref{VeryShort}, we obtain 
\begin{equation}
 \frac{c_\text{HR}}{96\pi r_h^4(v_0)V_b^2} 
 \ll - \frac{\kappa c\,r_h(v_b)}{r_h^4(v_0)(V_a-V_b)^3(U_a-U_b)} \ . 
 \label{VeryShort2}
\end{equation}
Thus, the quantum focusing condition is satisfied, 
\begin{equation}
 \frac{d\Theta}{d \lambda} < 0 \ . 
\end{equation}


\section{Conclusion and discussions}\label{sec:Conclusion}

In this paper, we have studied the quantum focusing conjecture for evaporating black holes. 
We have considered outgoing null surfaces near the event horizon of an evaporating black hole. 
We have simplified the model by using the s-wave approximation, and 
the matters are given by two-dimensional massless fields. 
Near the apparent horizon, the incoming energy is negative and outgoing energy is approximately zero. 
Then, the geometry near the event horizon is approximated by 
the ingoing Vaidya spacetime with the apparent horizon decreasing with time. 
The (classical) focusing condition is, in fact, violated near the apparent horizon. 

The quantum focusing condition states that the quantum expansion is non-increasing. 
The quantum expansion is defined by replacing the area in the (classical) expansion 
by the generalized entropy. 
The generalized entropy is the sum of the black hole entropy, 
which is given by the area of the horizon, 
and the entropy of matters outside the horizon. 
For the quantum expansion, the generalized entropy is further generalized 
to an arbitrary surface and given by the sum of the area of the surface and 
the entropy outside the surface. 
As the quantum expansion is defined by using the von Neumann entropy of 
the entropy of matters outside the surface, 
the generalized entropy is simply given by the entanglement entropy 
outside the surface including contributions from gravity. 
In this paper, we have considered outgoing null surfaces 
near the event horizon of the ingoing Vaidya spacetime, 
and the quantum expansion is given by the derivative (per unit area) of 
the entanglement entropy outside a cross-section, which we call $b$, of the null surface. 

We have calculated the entanglement entropy by using the island rule. 
The entanglement entropy effectively contains contributions from the island 
if the configuration with the island gives the minimum of the entanglement entropy 
in the saddle points in the replica trick. 
We have found that the Page time is given by an approximately null surface 
if it is defined by positions of $b$ where the transition 
from the configuration without islands to the configuration with the island occurs. 
If the (codimention-2) surface $b$ moves along a (codimension-1) timelike surface 
at a fixed distance from the event horizon, 
the surface $b$ passes across the Page time,
and the entanglement entropy follows the Page curve. 
However, if the surface $b$ moves along an outgoing null surface, 
it cannot intersect the Page time as the Page time is also an outgoing null surface. 
Thus, an outgoing null surface is either before or after the Page time, 
but cannot lie across the Page time. 

We have also found that the entanglement entropy is always increasing with time 
along outgoing null surfaces even if the outgoing null surface is after the Page time. 
The classical focusing condition is violated because the expansion is negative inside the apparent horizon. 
The Page curve implies that the entanglement entropy is decreasing with time after the Page time, 
and in fact, this behavior is reproduced if the surface $b$ moves along a timelike surface. 
One may naively expect that the quantum focusing conjecture is also violated 
because the entanglement entropy is decreasing after the Page time. 
However, the entanglement entropy is always increasing along outgoing null surfaces, 
or equivalently, the quantum expansion is always positive. 
Thus, the quantum focusing conjecture would not be violated even after the Page time. 

Unfortunately, the island rule possibly breaks down 
--- the saddle point for the quantum extremal surface disappears ---
if the surface $b$ is too close to the quantum extremal surface $a$. 
The (classical) expansion becomes negative inside the apparent horizon \eqref{AH}, 
and the apparent horizon is located inside the region where 
the saddle point of the quantum extremal surface disappears \eqref{crit-b}. 
Therefore, it is still unclear whether the quantum focusing conjecture 
is valid even in the region where the (classical) expansion is negative. 
However, the classical focusing condition is violated in a wider region \eqref{FTF}, 
and hence, there is a region where 
the quantum focusing condition is violated but the island rule is valid. 
We have calculated directly the derivative of the quantum expansion 
and seen that the quantum expansion is non-increasing 
even in the region where the classical focusing condition is violated. 

A most important feature of the entanglement entropy for the quantum focusing conjecture 
would be that it is still increasing with time even after the Page time. 
For timelike surfaces, outgoing modes pass through the timelike surface 
and moves from the inside to the outside of the surface. 
More and more information moves from the black hole side to the Hawking radiation, 
and then, the entanglement entropy starts decreasing after half of 
the information of the total system has moved to the Hawking radiation. 
For outgoing null surfaces, outgoing modes cannot go across the null surface, 
and the Hawking radiation outside the surface cannot get more information from the black hole. 
Thus, the entanglement entropy does not decrease even after the Page time, 
in contrast to the conventional setup for the Page curve. 

In this paper, we have explicitly shown that 
the quantum focusing conjecture holds even after the Page time. 
However, we have not provided a proof of the quantum focusing conjecture, 
but checked only a simplest but most interesting case of evaporating black holes. 
Moreover, we have studied mainly by using the s-wave approximation. 
Our calculation without using the s-wave approximation gives an exact result 
only if $a$ and $b$ is sufficiently close to each other, 
and we have made only a rough estimation for a larger distance 
by using some assumptions which are reasonable but to be justified. 
We do not pursue this direction and a complete check 
beyond the s-wave approximation is left for future studies. 
It would also be interesting to see the quantum focusing conjecture in more general cases.

\section*{Acknowledgments}

The author would like to thank Akihiro~Ishibashi for fruitful discussions. 
This work is supported in part by JSPS KAKENHI Grant No.~JP20K03930 and JP21H05186.

\begin{appendix}


\section*{Appendix}


\section{Entanglement entropy and vacuum states}\label{app:entropy}

Here, we briefly review the formula of the entanglement entropy. 
The entanglement entropy of a region can be calculated by using the replica trick 
\cite{Callan:1994py,Holzhey:1994we,Calabrese:2004eu,Casini:2005rm,Calabrese:2009qy}, 
\begin{equation}
 S = \lim_{n\to 1} \frac{1}{1-n}\log \frac{Z_n}{Z_1^n} \ , 
 \label{replica}
\end{equation}
where $Z_n$ is the partition function on $n$-replica spacetime 
which consists of $n$ copies of the original spacetime, 
but the branch cuts are inserted on each connected part of the region.%
\footnote{%
In this paper, we focus on spherically symmetric configurations 
and consider the two-dimensional spacetime by integrating out the angular directions. 
} 
For the gravity part, we just substitute the classical solution, 
and $Z_n$ and $Z_1^n$ cancel with each other except for 
the contribution from the conical singularity at the branch points, 
which gives the area of the branch point \cite{Lewkowycz:2013nqa,Faulkner:2013ana}. 
Thus, the first term in \eqref{S(R)} comes from the gravity part.%
\footnote{%
The branch cuts are inserted both on the region $R$ and island $I$, 
and hence, the area term appears both for $a$ and $b$, 
which are boundaries of $I$ and $R$, respectively. 
Note that the area term for $b$ is absent in models without gravity around $b$ 
but is present in our setup. 
} 

The second term of \eqref{S(R)} is contributions from matter fields and given by 
the correlation function of twist operators 
which are inserted at the branch points. 
When branch cuts are located in $(x_i, x_i')$, $i = 1,2,\cdots$, 
the entanglement entropy is formally expressed as 
\begin{equation}
 S 
 = 
 \frac{c}{6} \sum_{i,j}\log\left|x_i-x'_j\right|^2 
 - \frac{c}{12} \sum_{i,j}\log\left|x_i-x_j\right|^2 
 - \frac{c}{12} \sum_{i,j}\log\left|x'_i-x'_j\right|^2 \ , 
 \label{Smatter0}
\end{equation}
where in terms of the null coordinates $(u,v)$, 
\begin{equation}
 |x_i-x_j|^2 = \left|(u_i-u_j)(v_i-v_j)\right| \ , 
\end{equation}
and similarly for $|x_i-x'_j|$ and $|x'_i-x'_j|$. 
The expression \eqref{Smatter0} includes self-interactions of the twist operators for $i=j$ 
and gives the UV divergence, which should be regularized by introducing a cutoff, 
\begin{align}
 u_i-u_i &\to \epsilon^u \ , 
 &
 v_i-v_i &\to \epsilon^v \ .  
\end{align}
For curved spacetimes, the cutoff should be given in a proper distance, 
\begin{equation}
 \epsilon^2 = - 2 g_{uv} \epsilon^u \epsilon^v \ . 
\end{equation}
Then, the entanglement entropy is expressed as 
\begin{align}
 S 
 &= 
 \frac{c}{6} \sum_{i,j}\log\left|x_i-x'_j\right|^2 
 - \frac{c}{6} \sum_{i<j}\log\left|x_i-x_j\right|^2 
 - \frac{c}{6} \sum_{i<j}\log\left|x'_i-x'_j\right|^2 
\notag\\&\quad 
 + \frac{c}{12} \sum_i \left|\frac{4g_{uv}(x_i)g_{uv}(x'_i)}{\epsilon^4}\right| \ . 
\end{align}
We ignore the constant terms, and then, the entanglement entropy of matters is given by \eqref{Smatter}. 
The expression above obviously depends on coordinates and 
is valid only for an appropriate choice of coordinates. 
The entanglement entropy depends on the quantum state of matter fields 
and the replica trick is applicable only for special states such as vacuum states. 
In order to find suitable coordinates for the expression \eqref{Smatter}, 
we consider the correlation function of the original massless fields 
since twist operators are composite operators of the original fields. 

For example, we consider a massless scalar field in the two dimensional spacetime, 
whose action is given by 
\begin{equation}
 \mathcal I = - \frac{1}{2} \int d^2 x \sqrt{-g}\, g^{\mu\nu} \partial_\mu \phi \partial_\nu \phi 
 = - \frac{1}{2} \int du\,dv\ \partial_u \phi \partial_v \phi \ , 
\end{equation}
where $u$ and $v$ are the null coordinates. 
The scalar field $\phi$ is expanded in the Fourier modes as 
\begin{equation}
 \phi(x) 
 = 
 \int_0^\infty \frac{ d \omega}{2\pi} \frac{1}{\sqrt{2\omega}} 
 \left[a_\omega e^{-i\omega v} + a_\omega^\dag e^{i\omega v} 
 + b_\omega e^{-i\omega u} + b_\omega^\dag e^{i\omega u}\right] \ , 
\end{equation}
where $a_\omega$, $a_\omega^\dag$, $b_\omega$ and $b_\omega^\dag$ 
are annihilation and creation operators of incoming and outgoing modes, respectively. 
The vacuum state $\left|0\right\rangle$ is annihilated by the annihilation operators, 
\begin{align}
 a_\omega |0\rangle &= 0 \ , 
 &
 b_\omega |0\rangle &= 0 \ , 
\end{align}
and the two-point correlation function for the vacuum state $\left|0\right\rangle$ 
is calculated as 
\begin{align}
 \langle 0 | \phi(x) \phi(x') |0\rangle 
 &= 
 - \frac{1}{4\pi}\log \left|(u-u')(v-v')\right| \ . 
 \label{2pt0}
\end{align}
The energy-momentum tensor $T_{\mu\nu}$ is a composite operator of $\phi$, 
but a naive expectation value of classical expression has divergence, for example, 
\begin{equation}
 \lim_{x'\to x} \left\langle\partial_u \phi(x) \partial_{u'} \phi(x')\right\rangle 
 = - \infty \ . 
\end{equation}
Hence, we need to introduce the regularization to define the energy-momentum tensor. 
In the flat spacetime, 
\begin{equation}
 ds^2 = - du\, dv \ , 
\end{equation}
the energy-momentum tensor is regularized by using the normal ordering, 
\begin{align}
 \left\langle\,:\!\phi(x)\phi(x')\!:\,\right\rangle 
 &= 
 \left\langle\phi(x)\phi(x')\right\rangle 
 + \int \frac{ d \omega}{4\pi \omega} 
 \left\{[a_\omega,a_\omega^\dag]\, e^{-i\omega(v-v')} 
 + [b_\omega,b_\omega^\dag]\, e^{-i\omega(u-u')}\right\}
 \notag\\
 &= 
 \left\langle\phi(x)\phi(x')\right\rangle 
 + \frac{1}{4\pi}\log \left|(u-u')(v-v')\right| \ . 
 \label{NO}
\end{align}
Then, the energy-momentum tensor is zero for the vacuum state, 
\begin{equation}
 \langle 0 | T_{\mu\nu} |0\rangle = 0 \ . 
\end{equation}

Since the action of the two-dimensional massless scalar does not depend on the metric, 
the Fourier expansion of the scalar field can be given in arbitrary null coordinates. 
We define a new pair of the annihilation and creation operators by 
the Fourier modes in the coordinates $U$ and $V$ as 
\begin{equation}
 \phi(x) 
 = 
 \int_0^\infty \frac{ d \omega}{2\pi} \frac{1}{\sqrt{2\omega}} 
 \left[\tilde a_\omega e^{-i\omega V} + \tilde a_\omega^\dag e^{i\omega V} 
 + \tilde b_\omega e^{-i\omega U} + \tilde b_\omega^\dag e^{i\omega U}\right] \ , 
\end{equation}
and then, the vacuum state $|\Omega\rangle$, which is associated to $U$ and $V$, is defined by 
\begin{align}
 \tilde a_\omega |\Omega\rangle &= 0 \ , 
 &
 \tilde b_\omega |\Omega\rangle &= 0 \ . 
\end{align}
The two-point correlation function for the vacuum state $\left|\Omega\right\rangle$ is 
\begin{align}
 \langle\Omega| \phi(x) \phi(x') |\Omega\rangle 
 &= 
 - \frac{1}{4\pi}\log \left|(U-U')(V-V')\right| 
\notag\\
 &= 
 - \frac{1}{4\pi}\log \left|(u-u')(v-v')\right| 
 + G(v,v') + \bar G(u,u') 
 \ , 
 \label{2ptO}
\end{align}
where $G(v,v')$ and $\bar G(u,u')$ are regular functions. 
The energy-momentum tensor which is defined by using the normal ordering \eqref{NO} 
is non-zero for the vacuum state $|\Omega\rangle$, for example, 
\begin{align}
 \langle\Omega| T_{uu} |\Omega\rangle 
 &= - \frac{1}{4\pi} \lim_{u'\to u} 
 \partial_u \partial_{u'} \log\left|\frac{U(u)-U(u')}{u-u'}\right|  
 \notag\\
 &= - \frac{1}{24\pi} \left\{U,u\right\} \ , 
 \label{TuuSD}
\end{align}
where $\{f,x\}$ is the Schwarzian derivative. 

Thus, the correlation function is expressed in terms of 
the null coordinates of the flat spacetime as \eqref{2pt0} 
if the energy-momentum tensor is zero, 
but has additional regular terms as \eqref{2ptO} if the energy-momentum tensor is non-zero. 
In the case of the evaporating black hole, 
the energy-momentum tensor \eqref{TuuSD} gives nothing but 
the outgoing energy of the Hawking radiation, 
where $U$ is the retarded time before the gravitational collapse 
and $u$ is the standard null coordinate of the flat spacetime around the future null infinity. 
Since the energy-momentum tensor $T_{uu}$ is non-zero in the future null infinity, 
the two-point correlation function should be expressed in terms of $U$, 
which is different from the standard null coordinate in the future null infinity, $u$. 
For incoming modes, the energy-momentum tensor is zero in the past null infinity, 
and hence, the correlation function is expressed by using $v$. 
Since the twist operator is also a composite operator of the field, 
the expression \eqref{Smatter} should be expressed in terms of the same coordinates.

\end{appendix}

\end{document}